\documentclass[acmlarge]{acmart}
\AtBeginDocument{\settopmatter{printacmref=true}}
\usepackage[table]{xcolor}
\usepackage[disable]{todonotes}
\usepackage{natbib}
\usepackage{algorithm}
\usepackage{algorithmic}
\usepackage{multirow}
\usepackage{booktabs}
\usepackage{float}
\usepackage[inline]{enumitem}
\usepackage{caption}
\usepackage{subcaption}
\usepackage{cleveref}
\def\code#1{\texttt{#1}}

\usepackage[suppress]{color-edits} %

\makeatletter
\newcommand{\myconfshort}{\acmConference@shortname}
\newcommand{\myconffull}{\acmConference@name}
\newcommand{\myconfdate}{\acmConference@date}
\newcommand{\myconfloc}{\acmConference@venue}
\AtBeginDocument{
  \fancypagestyle{firstpagestyle}{
\fancyhead{}%
\fancyfoot[C]{}%
  }
  \fancyhf{}
  \fancyhead[LO]{\@headfootfont\shorttitle}%
  \fancyhead[RE]{\@headfootfont\@shortauthors}%
  \fancyhead[LE]{\@headfootfont\footnotesize \myconfshort, \myconfdate, \myconfloc}%
  \fancyhead[RO]{\@headfootfont\footnotesize \myconfshort, \myconfdate, \myconfloc}%
  \fancyfoot[C]{}%
}
\makeatother
\acmBooktitle{\conffull\@ (\confshort), \confdate, \confloc}

\addauthor{ZT}{olive}
\addauthor{CRZ}{cyan}
\addauthor{RA}{yellow}

\everypar{\looseness=-1}

\makeatletter

\newcommand{\note}[4][]{} 
\everypar{\looseness=-1}

\newcommand{\eg}{\emph{e.g.}}

\AtBeginDocument{%
  }

\setcopyright{acmlicensed}
\copyrightyear{2018}
\acmYear{2018}
\acmDOI{XXXXXXX.XXXXXXX}

\acmJournal{POMACS}
\acmVolume{37}
\acmNumber{4}
\acmArticle{111}
\acmMonth{8}

\copyrightyear{2026}
\acmYear{2026}
\setcopyright{cc}
\setcctype{by-nc-nd}
\acmConference[FAccT '26]{The 2026 ACM Conference on Fairness, Accountability, and Transparency}{June 25--28, 2026}{Montreal, QC, Canada}
\acmBooktitle{The 2026 ACM Conference on Fairness, Accountability, and Transparency (FAccT '26), June 25--28, 2026, Montreal, QC, Canada}
\acmDOI{10.1145/3805689.3806740}
\acmISBN{979-8-4007-2596-8/2026/06}
\begin{document}

\title[Big AI's Regulatory Capture]{Big AI's Regulatory Capture: Mapping Industry Interference and Government Complicity}
\author{Abeba Birhane}
\orcid{0000-0001-6319-7937}

\affiliation{%
  \institution{AI Accountability Lab (AIAL), School of Computer Science and Statistics (SCSS), Trinity College Dublin}
  \city{Dublin}
  \country{Ireland}
}
\email{aial@tcd.ie}

\author{Riccardo Angius}
\orcid{0000-0003-0291-3332}
\affiliation{%
  \institution{AIAL, SCSS, Trinity College Dublin}
  \city{Dublin}
  \country{Ireland}
  }

\author{William Agnew}
\orcid{0000-0002-1362-554X}
\affiliation{%
  \institution{Human Computer Interaction Institute, Carnegie Mellon University}
  \city{Pittsburgh}
  \country{USA}
}

\author{Harshvardhan J. Pandit}
\orcid{0000-0002-5068-3714}
\affiliation{%
  \institution{AIAL, SCSS, Trinity College Dublin}
  \city{Dublin}
  \country{Ireland}
  }
\email{me@harshp.com}
  
\author{Bhaskar Mitra}
\orcid{0000-0002-5270-5550}
\affiliation{%
  \institution{Independent Researcher}
  \city{Tiohtià:ke / Montréal}
  \country{Canada}
}
\email{bhaskar.mitra@acm.org}

\author{Roel Dobbe}
\orcid{0000-0003-4633-7023} 
\affiliation{%
  \institution{Faculty of Technology, Policy and Management, Delft University of Technology} 
  \city{Delft}
  \country{Netherlands}}
\email{r.i.j.dobbe@tudelft.nl}

\author{Zeerak Talat}
\orcid{0000-0001-5503-867X}
\affiliation{%
  \institution{School of Informatics, University of Edinburgh
  \city{Edinburgh}
  \country{Scotland}}}
\email{z@zeerak.org}

\renewcommand{\shortauthors}{Birhane et al.}

\begin{abstract}

Over the past decade, the AI industry has come to exert an unprecedented economic, political and societal power and influence. The well-functioning of regulatory and oversight structures and processes that govern the industry thus have paramount ramifications for everything from fostering public trust in systems marketed as AI, the credibility of scientific knowledge, educational and healthcare services and products, information ecosystems, %
the environment, rule of law and integrity of democratic process. %
It it therefore critical that we comprehend the extent and depth of pervasive and multifaceted capture of AI regulation {by corporate actors} in order to contend and challenge it. In this paper, we first develop a taxonomy of mechanisms enabling capture to provide a comprehensive understanding of the problem. %
Grounded in %
design science research (DSR) methodologies and extensive scoping review of %
existing literature and media reports,  %
our taxonomy of capture consists of 27 mechanisms across five categories.  %
We then develop an annotation template incorporating our taxonomy, and %
manually annotate and analyse 100 news articles. The purpose behind this analysis is twofold: validate our taxonomy and provide a novel quantification of capture mechanisms and dominant narratives. Our analysis identifies 249 instances of capture mechanisms%
, often %
co-occurring  with narratives that rationalise such capture.  We find that the most recurring categories of mechanisms are \emph{Discourse \& Epistemic Influence}, concerning narrative framing,  and \emph{Elusion of law}, related to violations and contentious interpretations of antitrust, privacy, copyright and labour laws. 
We further find %
that 
\textit{Regulation stifles innovation}, \textit{Red tape} and \textit{National Interest} are the most frequently invoked %
narratives used to rationalise capture.  
We emphasize the extent and breadth of regulatory capture by coalescing forces --- Big AI and governments --- as something policy makers and the public ought to treat as an emergency. Finally, we put forward %
key lessons learned from other industries along with
transferable tactics %
for uncovering, resisting and challenging Big AI capture as well as in envisioning counter narratives. %

\end{abstract}

\maketitle

\section{Introduction}
\label{sec:introduction}

The past decade has seen a rapid growth in the development and integration of AI technologies %
altering virtually all %
societal infrastructures -- such as information ecosystems (e.g., search and archives), education, finance, healthcare, and law enforcement -- %
affecting millions of people worldwide~\cite{de2024inescapable,ainowinstituteArtificialPower}. 
This has raised questions around intellectual property~\cite{pasquale2024consent,quintais2025generative}, democratic processes such as elections~\cite{olanipekun2025computational,nie2024artificial}, consumer protection~\cite{europaAntitrustCommission,cnbcAmazonWith,pandit2026terms}, algorithmic bias and discrimination~\cite{solaiman2023evaluating,dobbe2022system}, privacy~\cite{kalluri2025computer,feldstein2019global}, and misinformation~\cite{singh2025epistemic,hilal2024misinformation}. %
These concerns have in turn led to numerous global, multilateral, and national governance and regulatory efforts to preserve fundamental rights and to reduce the harms and risks of AI systems.
For example, the EU AI Act 
states its purpose is to ``\textit{promote the uptake of human-centric and trustworthy artificial intelligence (AI), while ensuring a high level of protection of health, safety, fundamental rights [...] including democracy, the rule of law and environmental protection, against the harmful effects of AI systems}''~\cite{AIAct}.
These aims reflect the expressed wishes of the public, which tend to be at odds with the controversial practices of  Big AI.\footnote{{We use the term `\textbf{Big AI}' to refer to the handful %
of companies that develop and mass deploy large-scale AI technologies -- such as large pre-trained models -- built on massive datasets collected through vast, centralised infrastructures, which, with increased integration into societal infrastructure, continue to exert outsized epistemic, economic, political, and societal influence. The term ‘\textit{Big AI}’ also encapsulates the structural consolidation of AI technologies by Big Tech as their core value proposition, central to their
infrastructure, resources, and strategic investments~\cite{van2024big}. 
 While new entities such as OpenAI, Anthropic, DeepSeek and xAI are included in our definition due to their significant geo-political influence, `Big AI' also marks the shift of existing Big Tech companies -- Alphabet, Meta, Amazon, Microsoft, Apple, NVIDIA -- towards becoming ``AI-first’’ companies, further expanding their unprecedented power and influence across all economic sectors and aspects of public life.}
}
 Recent public surveys have found that there is broad support for regulation of AI. For example,
  an analysis of 2000 responses on AI deployment found concerns related to violation of privacy and civil rights were among the most pressing, with the majority (68\%) of respondents highlighting these risks~\cite{robles2025artificial}. 
Similarly, Pew Research found that 60\% of the US population would be uncomfortable with the use of AI in their health care services~\cite{tyson202360}, while  72\% of respondents of a UK survey expressed a need for stronger regulation in order to feel comfortable with AI technologies~\cite{ada25}.
 Despite the broad public support for protecting public interests through regulation of the AI industry, there are growing concerns about how the tech industry's outsized influence on policy-making may obstruct %
meaningful safeguards and public priorities. %

Although the European Union has relied on civil society participation to shape digital regulation~\cite{european_center_for_not-for-profit_law_ecnl_towards_2024}, numerous investigations %
show  %
an outsized influence from the tech %
industry over the development of regulatory standards ~\cite{corporate_europe_observatory_bias_2025}.
For example, %
reports have shown how the European Commission has uncritically adopted the industry's call to ``simplify'' the AI Act (%
alongside other digital regulation) even before it has beenfully implemented~\cite{european_commission_simpler_2025}.
These concerns are also felt in the US, where a coalition of tech, consumer protection, labour, economic and environmental justice, and civil society organizations launched a ``\textit{People’s AI Action Plan}''as a counter-weight to the government's %
industry-backed AI executive orders and agenda: ``\textit{The White House AI Action Plan is written by Big Tech interests invested in advancing AI that’s used on us, not by us. Today, we are reclaiming agency over the trajectory AI will take: it’s time for a People’s Action Plan for AI that puts the needs of everyday Americans over corporate profits}''~\cite{ai_now_institute_peoples_2025}.

In contemporary societies, regulatory oversight functions as a central mechanism for ensuring public health, safety, and environmental protection~\cite{shleifer2005understanding}.
Regulation governs the breadth of society: from means of transport to the buildings we live in, the appliances used in our kitchens, the food available on the market, and educational infrastructure; it also defines mechanisms for health and safety inspections and protections.
Protective and social regulation, in particular, aim to ``\textit{make our lives safer by eliminating or reducing risks or exposure to risks}''~\cite{levi2011handbook}.
Regulatory and enforcement mechanisms have often emerged along with other practices---such as scientific codes of conduct addressing conflicts of interest---in the aftermath of tragedies that resulted in harms, injury, and death.
For example, the Thalidomide tragedy in the 1960s resulted in birth defects and deaths of thousands due to insufficient standards for testing drugs prior to their release to the public~\cite{moro2017thalidomide}.
Similarly, the Pfizer trial of the experimental antibiotic trovafloxacin (Trovan) for cerebrospinal meningitis in Nigeria resulted in at least 11 deaths and severe side effects in several children, including severe liver damage and kidney failure~\cite{loewenberg2008drug, carr2003pfizer}. Subsequently, both Thalidomide and Trovan were withdrawn from the market and reformed drug safety governance with the Thalidomide tragedy now considered a watershed moment for pharmaceutical regulation~\cite{moro2017thalidomide}.
Such incidents highlight how profit motives, %
conflicts of interest, and corporate influence may overshadow the public interest in  scientific research --- as well as in the development, marketing, and governance of consumer products --- and result in tragedy.\looseness=-1

Understanding corporate influence on AI regulation requires examining research and reporting on the actions and positions of the %
tech industry, particularly, Big AI, as well as those of policy-makers and other relevant actors.
A growing body of evidence shows how \textit{the AI industry} has been attempting to undermine and resist regulation, oversight, and enforcement~\cite{corporateeuropeChallengeThee}, including through large scale lobbying~\cite{corporateeuropetrojan,substackRevealedShocking}; retaliating against whistleblowers~\cite{metachildsafety}, civil society groups, researchers and law-makers~\cite{h1bvisa,bbcSocialMedia}; the revolving door---e.g. the former French Secretary of State for Digital Transition, C{\'e}dric O, becoming a shareholder and advisor for Mistral~\cite{corporateeuropetrojan}; and political donations~\cite{issueoneTechCozies}. 
For instance, Amazon, Google, Meta, and Apple each contributed \$1 million dollars to Donald Trumps political campaign, while Elon Musk contributed \$250 million to pro-Trump groups prior to the 2024 US elections. 

Governments and regulatory authorities can %
also play a key role in %
undermining and minimising the effects of existing or emerging rules. 
For example, the EU Commission President Von der Leyen has called for \textit{deregulation}~\cite{europaSpeechPresident}; the General Purpose AI Code of Practice under the AI Act --- drafted with industry participation --- was diluted at each stage until protections for human rights were made optional~\cite{cabrera2025human}; and the UK AI Bill has been delayed; while German authorities have discussed withdrawing laws to maintain ``\textit{attractive[ness] to tech companies}''~\cite{parliamentnewsBritainDelays} and to ensure ``\textit{innovation-friendly}'' implementation of AI regulation~\cite{fragdenstaatKoalitionsverhandlungenCDUCSUSPD}. 
A similar trend is evident in the US: the current administration has halted enforcement and regulatory oversight into Meta's alleged improper use of user financial data obtained from third parties for advertising purposes~\cite{citizenTrumpHalting}; halted, dropped, or withdrawn enforcement of 100s of cases of alleged corporate wrongdoing~\cite{citizenDeletingTech}; and signed an excursive order banning state-level AI regulations~\citep{robins2025trump}.

Beyond the converging actions between regulators and the tech industry, 
 coordinated efforts that actively campaign to portray the AI industry positively while framing regulatory oversight as undesirable put meaningful governance and regulatory oversight at risk.%
For example, 
the new super PAC formed by Andreessen Horowitz, Greg Brockman, and Mark Zuckerberg has pledged tens of millions of dollars to promote ``industry-friendly policies'' and cast the AI industry in a favourable light~\cite{techcrunchSiliconValley,nytimesSiliconValley}. 
Indeed, several public campaigns are already underway. A recent open letter signed by 100s of CEOs presents a simplistic dichotomy of ``\textit{innovation or regulation}''~\cite{LetterAICertainty}, alongside media outreach, targeted advertising, and funding of public initiatives, such as Meta's multimillion-dollar ad campaign depicting data centres as a boon to agricultural towns in Iowa and New Mexico~\cite{datacentre}.

The coalescence of corporate and state interests threatens the rule of law, independent regulatory oversight, and meaningful accountability --- all of which are crucial for the responsible and publicly supported technological innovation. 
It is therefore important to shed light on the various forms of corporate capture, the mechanisms involved, %
and the narratives used to legitimise it %
so that these practices and responsible actors can be better understood and challenged.
We interpret this as a reflexive knowledge problem; the growing role of capture in AI regulation and governance --- and its societal consequences --- is  a relevant but understudied %
topic in its own right and a factor that %
reflexively shapes the scholarship undertaken in FAccT and adjacent communities, e.g., by %
influencing narratives, research funding, infrastructure and priorities~\cite{whittaker2021steep,abdalla2021grey,young2022confronting}.

In this paper, we study %
mechanisms that enable capture of %
AI regulation, oversight, and public discourse. 
{
Based on a comprehensive study of grey %
and academic literature across different disciplines and societal actors working to uncover corporate capture (see Section~\ref{sec:related_work}), we identified (1) that corporate capture is a multi-faceted topic with few general \emph{conceptual frameworks} available,  (2) there is little academic research %
on %
corporate capture by Big %
AI %
specifically, and (3) %
existing work on corporate capture of AI regulation is mostly done by advocacy groups and civil society organizations, %
 but this has not been brought together to provide aholistic picture of the phenomenon. %
We address this gap by mapping  %
the %
types of mechanisms that enable capture and associated narratives as  this is an important %
step %
towards comprehensively %
understanding, identifying, documenting, and challenging capture as well as charting alternative futures.
We approached this knowledge problem by drawing on Design Science Research (DSR), and %
develop %
a taxonomy that serves as a %
\emph{conceptual model} for mapping capture mechanisms and narratives.}
We draw on desk research, expert exchange, literature review, and iterative design cycles in which we apply and evaluate the taxonomy and refine its comprehensiveness and clarity. %
Based on our taxonomy, we then develop an annotation schema %
and manually annotate two datasets: \texttt{DS1}, containing 25 articles and \texttt{DS2}, containing 75 Reuters news articles. The purpose of this analysis is twofold: the analysis of \texttt{DS1} validates our taxonomy, while analysis of \texttt{DS2} provides a novel quantification of capture mechanisms and narratives as well as insights into their prevalence and characteristics. 
{We emphasize here that our study is primarily \textit{descriptive and interpretive} %
and does not 
seek to quantify the extent to which capture occurs or mechanisms and narratives that cause it. 
}

We find that the landscape of capture is multi-faceted. 
Tech companies employ a wide variety of tactics: from  disregarding or misinterpreting existing laws; claiming %
regulation stifles innovationand unfairly prevents people from accessing AI products; relying on speculative studies; lobbying; %
to private meetings with regulators to ensure that regulation remains favourable and oversight negligent. 
{A}ddressing the current challenge of capture in the AI landscape requires %
drawing inspiration from past efforts against regulatory capture. %
We  therefore draw on past %
efforts from other sectors to counter regulatory capture %
as a guide for protecting %
public values that are vital to the present and future of our academic fields, public institutions, regulatory bodies, and shared public life.\looseness=-1

\section{Related work: %
the multifaceted nature of capture}
\label{sec:related_work}

Despite the rapidly expanding power, influence, and impacts on broad swathes of society, corporate capture in the AI domain remains poorly defined and understudied. 
Existing academic scholarship has examined forms of capture in other sectors, such as  %
the tobacco, oil, pharmaceutical, and information technology industries; however,  
a systematic examination and organisation of such work within the AI industry remains limited. 
This section addresses this gap by surveying disparate strands of relevant work from civil society organisations and academic literature. In accordance with  \citet{shapiro_blowout_2012}, we organise prior work into two functionally different forms of capture: %
\textit{Epistemic capture} and \textit{Regulatory capture}. %
\textit{Epistemic capture} can further be distinguished into \textit{Academic capture}  --- concerning the industry's influence over scientific knowledge production through, e.g., funding, sponsorship, and institutional partnerships that shape academic research intersecting with AI, including but not limited to computer engineering, ethics, and safety --- and \textit{Media capture}  --- pertaining to efforts to sway media, public discourse, debates, and conceptions of AI through strategies such as narrative framing, deception, disinformation, and doublespeak. 
By structuring the literature {thus}, we highlightthe fragmented nature of existing work {and the} converging themes that emerge across it. 

\vspace{-0.2cm}
\subsection{Epistemic Capture}
\subsubsection{Academic Capture}
Corporate capture extends to scientific knowledge production~\citep{krimsky1985corporate, hiltner2024fossil, muttitt2004degrees, noor2024elite, abdalla2021grey} which may intensify industry influence on regulation by legitimizing industry narratives and discredit narratives unfavourable to corporate interests.
Industry funding, for example, deliberately shapes scientific conclusions to favour industry and undermine public welfare~\citep{orr2010merchants, baba2005legislating, cranor2008tobacco, krimsky2004science, sass2005vinyl, michaels2008doubt, abdalla2021grey}.
\citet{lachapelle2024academic} identify three contributing factors for academic capture:
\begin{enumerate*}[label=(\roman*)]
\item increasing financialisation of higher education,
\item growing industry influence, and
\item reticence of university employees to challenge the status quo.
\end{enumerate*}
\citet{hiltner2024fossil} identify several types of  industry ties to higher education in the fossil fuel research community, including: serving on academic boards, 
sponsorship and endowments, student recruitment, advising on courses, and leasing of University lands.

The AI industry has been particularly successful in establishing their dominating influence over AI research~\citep{whittaker2021steep, ahmed2023growing}, paralleling the U.S. military’s dominance over scientific research during the Cold War~\citep{whittaker2021steep}. 
The concentration of control over data, compute resources, expertise, and funding has erected barriers to conducting critical AI research without industry involvement~\citep{murgia2019ai} or the AI industry's premise.
\citet{whittaker2021steep} adds, ``[t]hese companies control the tooling, development environments, languages, and software that define the AI research process --- they make the water in which AI research swims''.
The AI industry has created myriads of schemes to draw academia closer to the companies, including supporting  dual-affiliation arrangements that allow scholars to draw high salaries from Big Tech while publishing their research under academic affiliations, industry-sponsored Ph.D. programs, and joint grant programs  %
which align with and elevate industry perspectives~\citep{whittaker2021steep}.
\citet{bak2025risks} identify several mechanisms by which AI companies skew research findings including selective publishing of internal research, biased study design,  funding of academic research, inhibiting independent research, and selective collaborations.

Big Tech is also a major sponsor for preeminent venues for research on algorithmic harms --- \eg, the ACM Conferences on Fairness, Accountability, and Transparency (ACM FAccT); 
 on AI, Ethics, and Society (AAAI/ACM AIES); and on Human Factors in Computing Systems (ACM CHI) --- with many conference organizers and participants being affiliated with and drawing salary from these corporations~\citep{young2022confronting}.
Such sponsorship serves to bolster the image of corporations as socially responsible; influence events, decisions, and research; and identify academics who can be leveraged~\cite{abdalla2021grey}. %
These schemes to influence academic knowledge production also resemble strategies previously operationalized by the likes of Big Tobacco, Big Pharma, and Big Oil~\citep{abdalla2021grey, young2022confronting}; and raises serious concerns about research integrity in the field of AI. 
\subsubsection{Media Capture}
Capture of media and public discourse~\citep{schiffrin2018introduction, enikolopov2015media, hurt2004turning, taft2024oil} is another avenue for corporations to promote favourable narratives, sway public discourse, and suppress critique.
\citet{stiglitz2017toward} identifies four forms of media capture: 
\begin{enumerate*}[label=(\roman*)]
\item capture by ownership,
\item capture through financial incentives,
\item capture by censorship, and lastly,
\item cognitive capture.
\end{enumerate*}
The coverage of AI in mainstream media is consistently aligned with Big Tech. 
Companies' announcements and claims about their products %
are often reproduced with little, if any, scrutiny %
while simultaneously downplaying the credibility and expertise of those outside the industry. 
 
In an analysis of $1,000$ media articles ``[c]orporate actors [were] by far the most frequently quoted by journalists covering AI, with no civil society voices featuring amongst the top-25 most-quoted people or organisations''~\cite{tanner2023reframing}. Similarly, the New York Times quoted representatives of commercial tech companies $67\%$ of the time while only $6\%$ of quotes were from  representatives of civil society~\cite{barakat2024selective}. 
Moreover, industry insiders were described as `experts', while civil society representatives were presented as `sceptics'. 
Big Tech also exerts its influence over media through ownership, funding of journalism, shaping state media regulation, targeting media (policy) research institutions and universities, and by providing platforms through which the media comes to reach their audiences. 
In doing so, Big Tech has created an infrastructure in which media and journalists help reproduce, disseminate, and normalize industry narratives, or face exclusion~\citep{robins2025how}.
\vspace{-0.2cm}
\subsection{Regulatory Capture}

Capture of regulation occurs when %
powerful industries %
influence, manipulate, or undermine the governance agencies and authorities that %
oversee them,  \textit{control} decisions made about the industry~\cite{dal2006regulatory,levi2011handbook},  or \textit{steer} regulations towards  benefit for the industry rather than (or at the cost of) public interest~\cite{carpenter2013preventing,li2023regulatory}. 
{\citet{shapiro_blowout_2012} defines it as ``occurring when [regulating] agencies consistently adopt regulatory policies favo[u]red by regulated entities'', which encompasses influence, manipulation, and otherwise interfering with governance processes, rule-making, and independent oversight.} 
When regulators and governments become too sympathetic to the problems of the corporations and industries that they %
regulate, regulatory bodies risk becoming too lenient and, by extension, captured~\cite{levi2011handbook}. 
Regulatory capture can use legal (e.g,. lobbying) or illegal (e.g., bribes) tactics and may serve mutual government-industry interests, e.g.,  through relying on invited feedback from industry actors at the expense of their own research~\cite{agrell2012rethinking}. 
Other industries which have attempted regulatory capture have used a broad set of tactics. 
For example, the tobacco industry uses information manipulation, legal preemption, threats, financial incentives, and policy substitution~\cite{savell2014does}; while the pharmaceutical industry relies on lobbying and political influencing, such as by %
leveraging relationships with drug approval agencies, revolving doors, and challenging regulation %
through marketing campaigns, funding think tanks, and patient advocacy groups%
~\cite{vertinsky2021pharmaceutical, morgan2019cost}. 
Regulatory capture can also take the form of developing policy based on industry perspectives or by allowing industry influence to set policy agendas, manipulate information, or directly influence political~\citep{estache2011anti} and other forms of decision-making processes~\citep{li2023regulatory}.\looseness=-1

AI companies are found to evade regulatory enforcement and pressure regulators to change policy, for example, by withholding their digital  services in relevant jurisdictions~\citep{lancieri2024ai} while at the same time establishing partnerships with government agencies to redesign digital public infrastructure, which foster dependencies and capture government agencies~\citep{baykurt2025gov}.
In particular, the European Union's technology bills, including those for AI, are the biggest target of corporate lobbying~\citep{cerulus2025ranked} with tech companies  recruiting  large internal policy teams~\citep{hall2025investing}. Documents obtained by Corporate Europe Observatory~\citep{schyns2023lobbying} demonstrate how ``[v]ia years of direct pressure, covert groups, tech-funded experts [...] %
tech companies have reduced safety obligations, sidelined human rights and anti-discrimination concerns, and secured regulatory carveouts for some of their key AI products''. 
\citet{gorwa2024platform} and \citet{wei2024ai} identify access lobbying, coalition building, stakeholder mobilisation, funding, agenda-setting, academic capture, information management, cultural capture through status, and media capture through freedom of information requests and interviews with policy experts, respectively. 

\section{Methodology}
\subsection{Design Science Research Approach}
\label{sec:dsr}
We approach the %
study of capture mechanisms and %
associated prevalent narratives %
surrounding capture of AI regulation, %
through the methodological lens of \textit{Design Science Research} (DSR). 
DSR is a problem-solving paradigm that seeks to enhance %
in-depth and situated knowledge via the %
development of \textit{artifacts}~\cite{vom_brocke_introduction_2020}, originally intended to solve identified problems in contexts in which (a) existing theory is insufficient to explain the phenomena under study and (b) in which various aspects (people, organizations, technologies, institutions or other factors) need to be studied empirically and holistically to be comprehensively understood~\cite{hevner_design_2004}.
The designed artifact typically aims to support the knowledge problem and may take various forms~\cite{offermann_artifact_2010,weigand_artifact_2021}, which do not need be clear upfront. 
We identified a need for a taxonomy of mechanisms enabling capture as it quickly became clear that regulatory capture is prominent in global AI discussions{, major multilateral events, as well as day-to-day information exchange such as news articles.
And while various advocacy groups and authorities have reported valuable insights on various (sets of) capture mechanisms (see Section~\ref{sec:related_work}), a clear taxonomy and conceptual interpretation of corporate capture across all known or reported mechanisms was lacking.}

A DSR project is embodied by three closely related cycles of activities of understanding the problem environment (\textit{relevance cycle}), grounding a problem in theory and concepts (\textit{rigour cycle}), and designing and evaluating the artifact to support a knowledge need or solve a particular problem (\textit{design cycle})~\cite{hevner_three_2007}.
These cycles may happen in arbitrary order based on the structure of the research project, occur sequentially or in parallel, and may be iteratively revisited.
Our design science research followed the following six cycles: \\
\textbf{Cycle (1) \textit{relevance}:} To gain an in-depth understanding of the prevalence of corporate capture, we engaged in desk research and expert exchange. 
A group of seven experts gathered on a weekly basis for over a period of 10 months and discussed regulatory capture based on news articles that had surfaced in the individuals news consumption, and which were selected based on two inclusion criteria:
Relevance--a report is %
about or directly related to AI regulation--and credibility--which captures the reputability of the publishing venue  and the clarity and quality of the reported evidence. 
 We excluded reports that were not directly related ot AI regulation, e.g., meta-analyses, opinion pieces, etc., and reports with unsubstantiated claims.
The notes taken during desk research and expert meetings formed the basis for a literature review and the initial taxonomy development. 
In this cycle, we also determined the criteria for evaluating the taxonomy:  \textit{Comprehensiveness}--do the identified capture mechanisms cover the phenomena reported in the news in an exhaustive, to the extent possible, and mutually exclusive manner?--and \textit{legibility}--are the mechanisms and their categories and definitions clear enough to be interpreted consistently across a group of experts?
{To address the potential of remaining confirmation bias in the selection of articles in this first cycle, we decided on mitigating measures for evaluation (random sampling, see Cycle 4) and application (a structured systematized search of news articles, see Cycle 5).}\\
\textbf{Cycle (2) \textit{rigour}:} To ensure a comprehensive understanding of corporate capture, we performed %
a focused reading %
on existing definitions from academic literature and on analyses of current occurrences of capture--primarily through policy reports and commentary analyses from civil society and rights groups.
Insights from this literature review are reported in Section~\ref{sec:related_work}. \\
\textbf{Cycle (3) \textit{design and evaluation}:} %
Based on the notes gathered from the expert exchange (Cycle 1) and the literature review (Cycle 2), we developed an initial taxonomy. 
This taxonomy included alist of categories of corporate capture andcapture mechanisms within these categories. 
We iterated on these following discussions in expert meetings. We added three additional categories (\emph{Government adopting industry framing},
\emph{Conflation of public and private interest}, and \emph{No capture}) to the initial taxonomy andannotation schema (see Cycle 4).\\ 
\textbf{Cycle (4) \textit{design and evaluation}:} To further evaluate the
 taxonomy, we translated it to an annotation schema (see~\Cref{appendix:codebook} for the annotation template)
which we applied to %
\texttt{DS1}, %
composed of 25 %
articles randomly drawn from an initial set of 100 news articles collected by experts. See Section~\ref{sec:annotation} for  annotation and coding procedures. \\ 
\textbf{Cycle (5) \textit{relevance - applying the taxonomy}:} We applied %
the updated taxonomy and annotation schema to \texttt{DS2}, a dataset constructed from a systematic search (see Section~\ref{sec:dataset2}). %
Further, we extended the annotation schema to extract narratives used to support corporate capture, as reported in the articles. 
 \\
\textbf{Cycle (6) \textit{design and evaluation - finalizing the taxonomy}:} The final taxonomy was informed by another abductive step that was based on analyses by the annotators in Cycle 5 and subsequent evaluation by the experts group, focused on comprehensiveness and legibility of the identified mechanisms, their categories, and definitions. We  validated the presence of mechanisms in the studied datasets \texttt{DS1} and \texttt{DS2} through quantitative analysis of mechanisms and narratives, reported in Section~\ref{sec:results}. We summarize the final taxonomy in Section~\ref{sec:taxonomy} with the full taxonomy and descriptions given in \Cref{appendix:taxonomy}.%

\vspace{-0.2cm}

\subsection{Dataset Curation and Search Criteria  %
}
\label{sec:dataset2}
To study regulatory capture as it unfolds across globally significant events, %
we conducted an in-depth scoping review of news media articles, as these seek to report current affairs and reflect key debates and decisions that are taking place in the real world pertaining to the AI industry. %
News articles also function as a mechanism for obtaining up-to-date information in a fast-paced and continually shifting landscape.
We curated two datasets: \texttt{DS1} (25 articles), used to validate  our taxonomy %
and \texttt{DS2} (75 articles), used to quantify and perform in-depth analysis of capture as {reported in Reuters coverage 
}. %

DS1 was curated from news articles shared by researchers in a privatediscussion forum.
\texttt{DS2} was curated by searchingGoogle News for articles published around three periods which contain four events critical to the development of AI regulation: The first global AI Summit in the U.K. and the EU AI Act trilogue negotiations (October 2023-February 2024), the second global AI Summit in South Korea (April 2024-June 2024), and the third global AI Summit in France (January 2025-March 2025).\footnote{We searched using the keywords ``artificial intelligence policy \{ usa | eu | uk \}, obtaining 10.691 results for the USA, 8.106 for the EU, and 5.832 for the U.K. queries, respectively.} 

Our search results for \texttt{DS2} comprised 24.629 article URL and title pairs,
which we scraped using a Crawlee-based scraper.\footnote{Crawlee: https://crawlee.dev.} 
We de-duplicatedand sub-selected articles {based on rated quality, reliability, and orientation towards fact-based reporting. 
 We selected Reuters, rated as the most reliable and high-quality source in an analysis of 11.520 news domains~\cite{lin_high_2023},  as our single source.
 }
We %
used the article titles to compile a list of relevant unigrams, and applied this to identify  titles containingrelevant terms, resulting in 683 unique articles.\footnote{We kept only articles containingat least one of the following words  in their title: \code{\{nvidia, openai, musk, google, meta, ceo, microsoft, apple, deepseek, amazon, summit, ai act, regulat*, antitrust, policy, plan, rule\}}.}

We then manually applied our inclusion criteria (\textit{relevance} and \textit{credibility}, see Section~\ref{sec:dsr}) selecting articles based on title, which reduced the article count to 181 unique articles. 
From this set, we sampled 75 articles with a stratified sampling across the three periods, resulting in 26 samples for the first period (UK AI Summit \& EU AI Act trilogue negotiations), 16 samples for the second period (South Korea AI summit), and 33 samples for the third period (France AI Action Summit). 
 We then manually checked the body text of the articles to remove and replace, while maintaining stratification, duplicates -- including articles with a high degree of content similarity -- and articles that did not cover AI regulations (despite their relevant titles).

\begin{figure}[hbt!]
\centering

\centering
\includegraphics[width=0.5\textwidth]{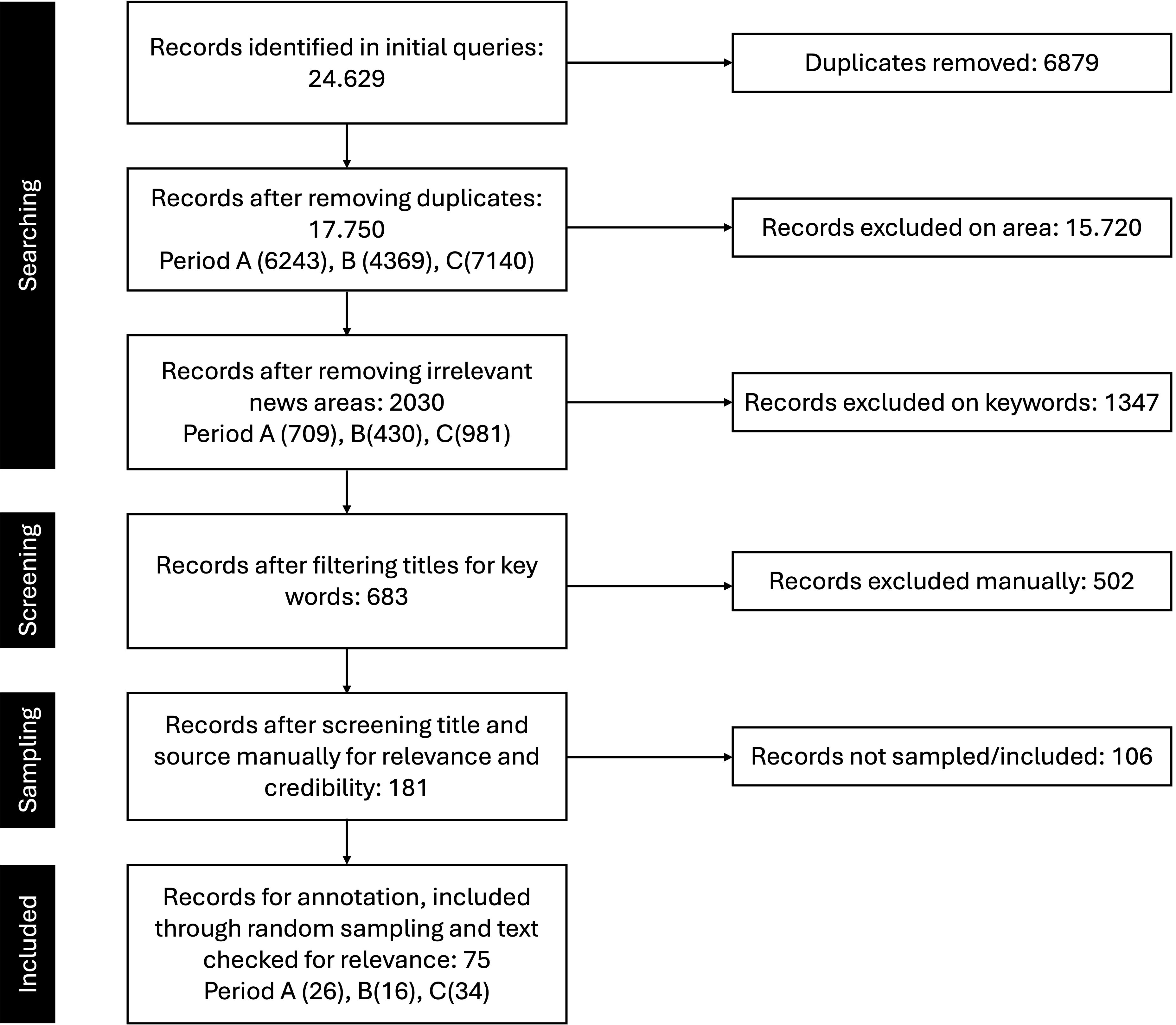}

\caption{PRISMA diagram for the construction of dataset \texttt{DS2}.}

\label{fig:prisma}

\end{figure}

\subsection{Annotation schema and code book}
\label{sec:annotation}
We developed the annotation schema iteratively informed by %
our taxonomy %
and expert discussions. %
The codebook (see \Cref{appendix:codebook}) contained fields recording article metadata (such as the publication date, title, data split -- i.e., \texttt{DS1} or \texttt{DS2} -- and time period), mechanism category, and mechanisms and narratives observed. 
Each article was annotated independently by two members of the research team. 
Annotators were instructed to carefully read each article and annotate: mechanism category, mechanism, narratives used, and excerpts containing the supporting evidence. %
The mechanisms  and their category were selected from our taxonomy (see \Cref{sec:taxonomy}), while narratives were given as a free-text field.
The iterativetaxonomy  development  allowed the team to consistently align with each other when identifying, documenting, and categorizing information. 
 To ensure a shared understanding of key concepts, we developed a common vocabulary of definitions, descriptions, and instructions as part of the code book.
Annotators extracted excerpts of the articles as `evidence' {for} their labels and to assist in later discussions and resolutions. 
After preliminary annotation, both annotators had labelled exactly the same mechanisms for 21 of the 100 articles,  44 had at least one  shared {mechanism,}  while 35 had no matching annotated mechanisms. 

We approach the annotation task as a consensus annotation task \cite{dehghan_dealing_2025}. 
Once independent annotation had been completed, annotators synchronously considered each disagreed-upon annotation and collectively decided on a final set of labels based on a manual review of the annotation and evidence collected by both annotators.\footnote{The deliberation processes occurred first synchronously to ensure a shared understanding, then asynchronously with one annotator raising instances for the other annotator to review and propose changes or accept the final label set.}

In line with \citet{oortwin2021}, this phase of disagreement resolution surfaced simple mistakes, due to missed evidence, interpretive disagreement {arising from} different interpretation of the same evidence, or conceptual misalignment related to misinterpreted taxonomy definitions.

Interpretive disagreement arose in two cases regarding breach of copyright and privacy laws, where both annotators befittingly attributed \textit{Disregard existing laws}, but only one attributed \textit{Misinterpret laws} to reflect that the company also publicly maintained they had operated within the bounds of law. The distinction between the two mechanisms was further illustrated by reports in which the latter was present, but the former was not.  This was the case for the reported roll-out of a personal data processing \emph{Consent or Pay} model. Here, the requirements of law were not disregarded altogether, yet misconstrued to mimic compliance. 

An example of conceptual misalignment encountered pertains to an article reporting %
a partnership between Mistral and the German defence startup Helsing~\cite{ExclusiveMistralSeeks} reported as an attempt from Mistral %
to build stronger ties with the German government. This was noted as \textit{Economic coercion of government} by one annotator but not the other. %
The mechanism was subsequently %
 clarified to encompass only explicit threats, and the annotation was dropped. We excluded mechanisms where  consensus could not be reached because the evidence was deemed insufficient by one of the annotators from the final labelled dataset.

\begin{table}[!htbp]
\centering
\caption{Taxonomy of capture mechanisms with brief descriptions. The highlighted concepts denote five broad, high-level categories, %
each further comprising a set of detailed mechanism categories and descriptions. %
}
\small
\begin{tabular}{|p{5.15cm}|p{10cm}|}
\hline
\rowcolor{gray!50}\textbf{Mechanism} & \textbf{Brief Description} \\ \hline
\rowcolor{gray!25}\textbf{Direct influence on policy} & \textbf{Influence position or decisions of public officials and regulations} \\ \hline
Lobbying & Communication intended to influence decisions or create favourable positions \\ \hline
Private meetings & Meetings or communications outside lobbying \& regulated channels \\ \hline
Political contributions & Contributions by a person or %
a company to a political entity or %
an authority\\ \hline
Economic coercion of government & Potential economic advantage or disadvantage as a threat to induce changes \\ \hline
\rowcolor{gray!25}\textbf{Conflicting involvement} & \textbf{Inherent conflicts due to involvement of an entity in a specific position} \\ \hline
Revolving door & Public officials taking up conflicting roles in private entities or vice-versa \\ \hline
Direct involvement in rulemaking & Involvement in developing law or policy without legal mandate or authorisation \\ \hline
Ownership/Stake in company & Direct/indirect ownership or stake in regulated organisations by public officials \\ \hline
\rowcolor{gray!25}\textbf{Market influence} & \textbf{Participation, co-operation, or coercion of market actors }\\ \hline
Standard setting in consortia & Consortiums develop and dictate practices without other relevant stakeholders \\ \hline
Corraling SMEs/orgs to oppose & Imploring, organising, or using SMEs to oppose a policy or position \\ \hline
Standard setting via monopoly & Using a position of monopoly or dominance to dictate practices \\ \hline
Economic coercion of competition & Threat of potential economic advantage or disadvantage to force competitors \\ \hline
\rowcolor{gray!25}\textbf{Elusion of law} & \textbf{Mechanisms directly or indirectly against the spirit or letter of the law} \\ \hline
Disregard existing laws & Disregard the requirements or process established in laws \\ \hline
Misinterpret laws & Using alternative or intentional misrepresentation to avoid requirements \\ \hline
Relocation of development and labour & Moving labour deployments or setups from one location to another \\ \hline
Exploit weak regulations/jurisdictions & Moving to jurisdictions with considerably weaker regulations or enforcement \\ \hline
Retaliation %
& Acting, suppressing, or retaliating against whistleblowers or regulators \\ \hline
Bribery & Solicitation, payment, or favour in exchange for official actions or decisions \\ \hline
\rowcolor{gray!25}\textbf{Epistemic \& Discourse Influence} & \textbf{Shape or control knowledge production process or public discourse %
} \\ \hline
Corporate sponsorship of events & Sponsorship of events to promote the %
company or imply support to principles \\ \hline
Funding/sponsor research \& education & Direct or indirect funding and sponsorship of research and education \\ \hline
Public facing campaign & Use of media %
campaigns against a specific person, topic, decision, or action \\ \hline
Hyping technologies & Promoting %
AI capabilities  without %
empirical evidence or conceptual rigour \\ \hline
Playing victim & Public complaints %
of being subjected to unfair rules \\ \hline
Undermining risks/harms & Disregarding, hiding, or minimising the potential of risks or harms \\ \hline
Speculative studies & Studies, analysis, or reports that do not follow scientific rigour or practices \\ \hline
Ethics washing & Imposing as guardian of %
ethics, transparency, safety without meaningful actions \\ \hline
Government adopting industry framing & Repeating or only considering industry positions while diminishing others \\ \hline
Conflation of public and private interest & Irrationally mixing or implying private interests as being public benefits \\ \hline
\end{tabular}
\label{table:taxonomy}
\end{table}
\section{%
Mechanisms of Capture and Insights from Quantitative analysis}
\label{sec:taxonomy}
 In this section, we  present our taxonomy of  capture mechanisms and  detailed  {analyses}  of \texttt{DS1} and \texttt{DS2}.%

\subsection{Corporate Capture: a Taxonomy of Mechanisms}

Our taxonomy was developed through the iterative design method (see Section \ref{sec:dsr}) and is informed by %
prior work on %
capture occurring in %
 sectors such as Big Tobacco, Big Oil, and Big Pharma (see Section~\ref{sec:related_work}). %
The taxonomy (see Table~\ref{table:taxonomy} and  Appendix~\ref{appendix:taxonomy} for an overview and an expanded description of the taxonomy) is comprised of \textit{27 different capture mechanisms across five categories}, which describe the variety of tactics and the sophistication of Big AI and the wider tech industry's corporate capture efforts.
We intend for the taxonomy to function as a `living resource' which will require iterative updating as the landscape and mechanisms of corporate capture change.\looseness=-1

\label{sec:results}

\begin{figure}[hbt!]
\centering
\begin{subfigure}[t]{1\textwidth}

\centering
\includegraphics[width=0.5\textwidth]{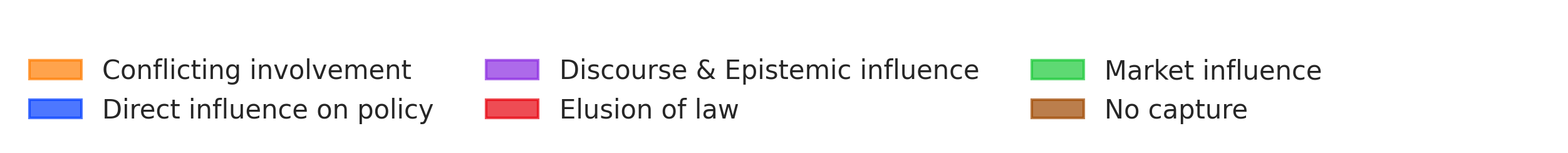}

\label{fig:corp-cap-legend}
\end{subfigure}
\begin{subfigure}[t]{.49\textwidth}\centering
\includegraphics[width=1\columnwidth]{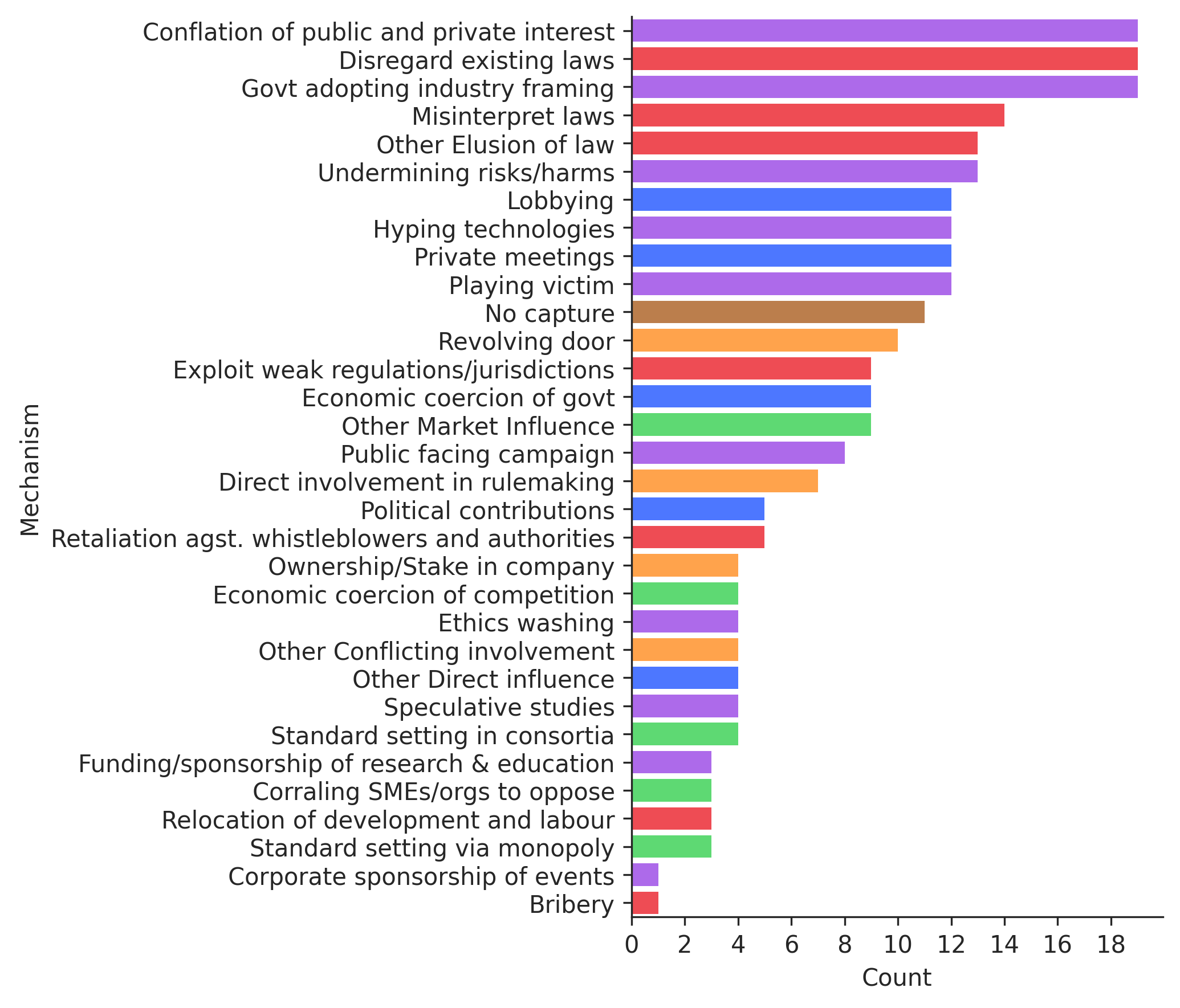}

\caption{Frequency of capture mechanisms in \texttt{DS1} \& \texttt{DS2}. \textit{Other} denotes when a broad category of mechanisms, but no specific mechanism, could be identified.
}
\label{fig:corp-cap-totals}
\end{subfigure}
\hfill
\begin{subfigure}[t]{.49\textwidth}\centering
\includegraphics[width=1\columnwidth]{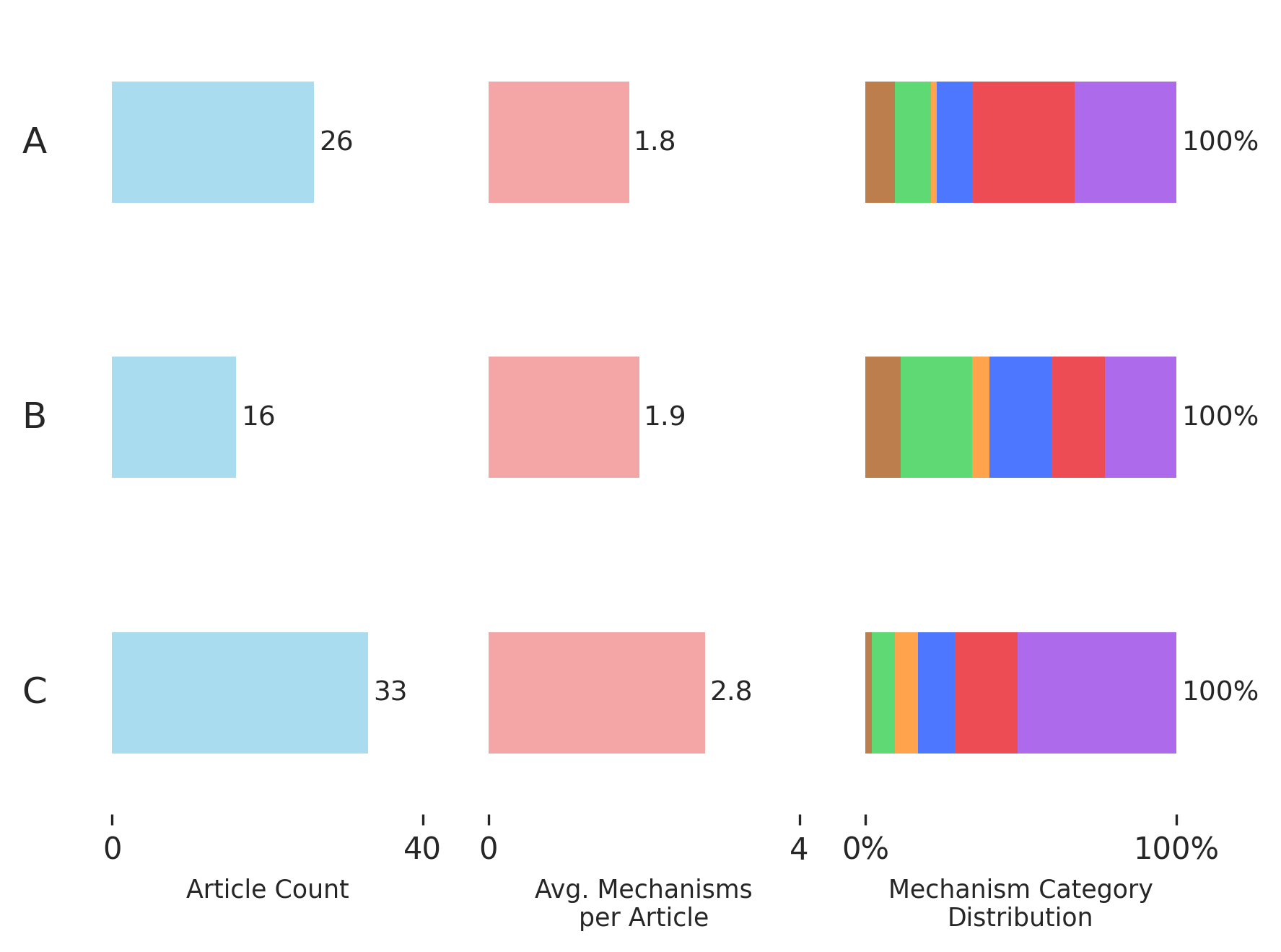}

\caption{
Article count, average mechanisms per article and mechanism category distribution for the 75 articles in \texttt{DS2} covering the three periods.
}

\label{fig:corp-cap-periods}

\end{subfigure}
\caption{Quantitative results for capture mechanisms found in the annotated datasets.} %

\label{fig:corp-cap-mech}
\end{figure}

\subsection{Capture mechanisms: distribution and relationships }

In Figure~\ref{fig:corp-cap-totals} we present the distribution of the 249 total instances of capture mechanisms: 79 (32\% of all instances) from %
\texttt{DS1} and 170 (68\%) from %
\texttt{DS2}.
We found capture mechanisms present in all but 11 articles from \texttt{DS2} -- labelled as \textit{no capture}. 

We found capture mechanisms in all articles, with the exception of 11 articles from \texttt{DS2} %
labelled as \textit{No capture}. This label was not applicable to any of the articles in \texttt{DS1}, as the original set used %
to form it was deliberately %
composed of articles that concerned both regulation and narrative capture.

The 10 most frequently identified mechanisms (50\% of instances) all belong to either \textit{Discourse \& Epistemic influence} (D\&EI), \textit{Elusion of law}, or \textit{Direct influence on policy} categories. Within this set, \textit{Elusion of law} is the most recurring category outside of narrative-framing activity, and comprises  violations - \textit{Disregard existing laws} (17\%) - and contentious interpretations - \textit{Misinterpret laws} (14\%) - of antitrust, privacy, copyright and labour laws, as well as other mechanisms of operation against the spirit or letter of the law.

We further found that \textit{Lobbying} was present in 40\% and 3\% of \texttt{DS1} and \texttt{DS2}, respectively. 
Similarly, \textit{Revolving Door} occurred in 24\% and 5\% of \texttt{DS1} and \texttt{DS2}. 
These 10 instances of national relevance and high-profile \textit{Revolving Door} were distributed between the US (6 cases), the UK (3 cases), and EU (France, 1 case). 
Of these 10 cases of conflicting interests, we find that 4 involve ongoing \textit{Ownership/Stake in company} by public officials during their appointment to public office and are evenly distributed across the US and UK. 
The remaining 125 instances belong to other, less occurring capture mechanisms.%

For \texttt{DS2}, we count reported mechanisms across the three periods shown in Figure~\ref{fig:corp-cap-periods}. 
Period A spans five months (late 2023 to early 2024, surrounding the UK AI Summit and EU AI Act trilogues), whereas  Period B and Period C %
comprise three months intervals (around the 2024 Korean AI Summit and the 2025 Paris AI Action Summit, respectively). 
The article sample size for each period was calibrated against the number of results returned by %
the search. %
We observe that the distribution of labelled categories for each article are stable for Period A and B, with an average of approximately two mechanisms labelled for each article. 
For Period C, we find an average of approximately 3 mechanisms labelled for each article.

Across \texttt{DS1} and \texttt{DS2}, we find 55 articles labelled for \textit{D\&EI}. 
For these cases, each instance is co-reported with an average of 1.20 non-\textit{D\&EI} mechanism for each \textit{D\&EI} mechanism (1.63 and 1.03 in \texttt{DS1} and \texttt{DS2}, respectively).

\subsection{Recurring narrative framings and co-occurrence with capture mechanisms}

\begin{figure}[hbt!]

\centering
\includegraphics[width=0.5\textwidth]{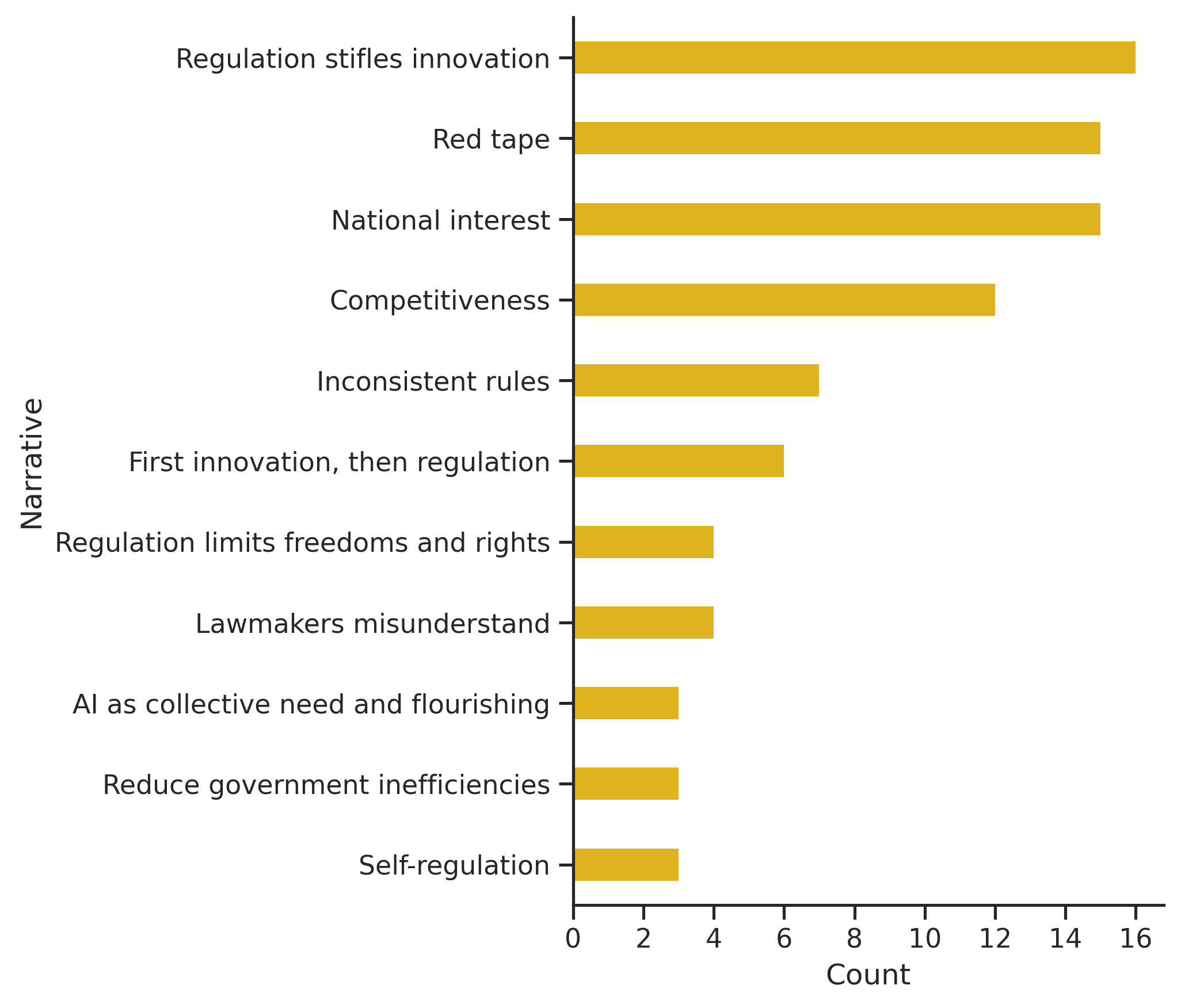}

\caption{Frequency of capture narratives across {\texttt{DS1} and \texttt{DS2}}.  }

\label{fig:corp-cap-narratives}
\end{figure}

Figure~\ref{fig:corp-cap-narratives} shows %
the total occurrences of capture narratives across annotated articles: %
49 out of the 100 articles %
contain %
narrative(s) that attempt to justify capture. 
We identify 11 recurring framings %
across the 49 (13 in \texttt{DS1}, 52\% of its total, and 36 in \texttt{DS2}, 48\%), {through a} manual thematic clustering of the free-text annotations of article excerpts {which reference the use of} narratives. 
For any {given} article, these clusters may appear alone or co-occur with others. %

\textit{Regulation stifles innovation.} We found this to be the most frequently occurring narrative (16\% overall; 24\% and 13\% in \texttt{DS1} and \texttt{DS2}, respectively), decrying regulation as  ontologically at odds with progress.  % 

\textit{Red tape.} The second most frequent narrative concerns alleged \textit{Red tape} with 15\% of all articles labelled for this narrative (28\% of \texttt{DS1} and 11\% of \texttt{DS2}). 
Within this narrative, regulation is portrayed as unnecessary, excessive, or obsolete, expressed through phrasing such as ``regulatory burden'', ``over-regulation'', ``simplification'', and ``cutting red tape''. %
We further observe that this framing tends to precede more explicit calls for ``deregulation'' from top-level regulatory authorities who have historically adopted such narratives
 and continue to do so within the sequence of arguments in their more recent speeches \cite{europaSpeechPresident}. %

This narrative co-occurs with mechanisms that imply direct contact with regulators -- 6 articles labelled with \textit{Lobbying} and 6 with \textit{Private meetings}. 

\textit{National interest.} This narrative aggregates calls against assumed impending threats to specific countries or geopolitical blocs. Examples include ``falling behind'' in technology development and  %
more explicit warnings concerning international economic leadership and national security, such as ``the AI race''.

\textit{Competitiveness.} Competitiveness was the fourth most frequently occurring framing (12\% overall, 24\% and 8\% in \texttt{DS1} and \texttt{DS2}, respectively).
This narrative category consolidates framings around the econometric parameters of productivity and competitiveness as values of higher priority than regulatory oversight. 
The \textit{Government adopting industry framing} mechanism appears most frequently together with the \textit{Competitiveness} narrative (eight articles), and to a lesser extent, with \textit{Regulation stifles innovation} and \textit{Red tape} narratives (five articles each).

\textit{Inconsistent rules} 
refers to the characterisation of existing regulation as unclear or fragmented. 
This line of argument occurs in our 7\% of our data (12\% of \texttt{DS1} and 5\% of \texttt{DS2}) and alleges both that there are difficulties in interpreting existing laws and that AI deployment must be globally uniform. 
Such narratives seek to portray regulation as time-intensive and unfeasible, and at odds with an increasingly globalised technical infrastructure. 
In one specific instance, localised regulations were self-contradictorily described as an obstacle to industry efforts to 
``accurately understand important regional languages, cultures or trending topics on social media'' \cite{bettya_building_2024}.

\textit{First innovation, then regulation.}
This family of low-frequency narratives argue that while regulation is needed, technological development outpaces regulatory innovation, and regulators should wait for technological infrastructure to mature into its full potential prior to regulation. 

Lawmakers are further cautioned against regulating technologies as \textit{Regulation limits freedoms and rights}--another low-frequency narrative occurring in our data which narrative contrasts regulation with ``potentially life-saving innovations'', and specific provisions for interoperability with ``user privacy and data security''. 
For instance, mandates to moderate and remove illegal content online are constructed as the legal ``institutionalisation of censorship''.
Relatedly, the \textit{Lawmakers misunderstand (technology)} narrative suggests that regulators lack the required expertise and understanding of technology to draft appropriate regulation, alleging that companies ``are better placed to uncover problems'' and that regulatory demands are ``unrealistic'' or ``unnecessary''.

Other, less prevalent supporting narratives include: characterising \textit{AI as a collective need and essential for flourishing} that should not be hindered by law; the neo-liberal aim to \textit{Reduce government inefficiencies}, which has  historically conflated all bureaucracy as unnecessary and inefficient bloat;
\textit{Self-regulation}, which acknowledges the need for rules, but privileges non-binding, industry-led pledges and codes of conduct.\looseness=-1

\section{Discussion}

\subsection{Summary of findings and implications}
The results in this paper provide %
{insight} into the variety and structure of mechanisms  in the analysed news articles.
While our results do not speak to potential correlations or causal effects of the observed mechanisms and regulatory outcomes, they can be used to inform hypotheses about the relationship between the 27 recurring capture mechanisms we have identified and regulatory outcomes. %
The extent %
of corporate capture further motivates the urgent need to characterize and address the growing centralization of power and influence in the hands of a few tech companies and their shareholders~\cite{ainowinstituteArtificialPower}, and its consequences for human rights~\cite{afp_concentration_2025}. 
Our work on capture mechanisms and narratives serves as an additional dimension in the growing chorus of investigations into the political economy of Big Tech (see also Section~\ref{sec:related_work}). %

{Prior} work has typically adopted a dichotomous theoretical lens -- %
distinguishing  information capture {from} influence on policymaking \cite{shapiro_blowout_2012}, whereas the most recurring mechanisms in our evidence also 
include %
the \textit{Elusion of law}. 
These recurring %
violations and contentious interpretations of antitrust, privacy, copyright and labour laws call into question the effectiveness of  enforcement.
Such pressures, along with weakening the mandates of regulatory agencies, risk the normalisation of a \textit{de facto} law in which Big AI operates outside the bounds of regulatory scrutiny. 
Furthermore, the stochastic nature of the underlying technology--where the technology cannot be reliably tested--coupled with the economic power of Big AI corporations enables a normalisation of ``algorithmic states of exception'' at scale.
Defined by \citet{mcquillan_resisting_2022}, this indicates the  %
application of algorithms %
as the de facto authority in contexts where their behaviour has not been or cannot be tested, creating conditions akin to martial law. 
The conjunction of  law-flouting %
practice and technological affordances thus raises urgent concerns regarding the contemporary integrity of lawmaking institutions %
over their respective %
jurisdictions.

Our annotation of narratives employed provide another qualitative avenue for insight into capture strategies, which can be further studied to understand the causal or system impact of different narratives on discourse, public conceptions of AI, and knowledge production. 
Our finding that there is substantial growth over time in the \textit{Discourse \& Epistemic Influence} (see Figure~\ref{fig:corp-cap-periods}) indicates that how AI is framed is becoming increasingly important.
The consistent co-occurrence of each D\&EI mechanism with approximately one corresponding mechanism from other categories indicates the significant role that public-facing campaigns play. 
Public facing campaigns often occurred in parallel with lower profile activity such as direct influence on policy, abuse of monopoly power, elusion of law and muddled roles between regulating agencies and the industries to be regulated.

Another core element in our findings is the \textit{central role played by governments and public officials} across a subset of mechanisms and narratives. 
While our analysis cannot speak to the impact of the regulators entanglement with corporate actors, it corroborates concerns about public institutions' opposition to regulation and outright efforts to reduce the application of existing regulation in the US, UK, and EU. 

Our observations {illustrate} the %
tolerance of  high-profile \textit{Revolving door}s, muddling roles between regulated and regulators, and more severe cases of active \emph{Ownership/Stake in Company} by incumbent public officials. 
Such entanglement erodes 
public trust %
in the capability and willingness of institutions to scrutinise corporations and enforce laws. %

Governments are deeply dependent on Big Tech infrastructure.
AI is broadly perceived by policy makers as central to the economic investments and growth plans of many countries.
However, there are growing concerns regarding a lack of return on investments in AI technologies across economies~\cite{the_economist_investors_2025} and %
circular investments in the AI industry are contributing to increasing systemic risks to the global economy~\cite{arun_bubble_2025}.
Of special concern are a cluster of narratives calling to \emph{Reduce government inefficiencies} through the wholesale replacement of civil servants with experimental AI technologies at the behest of private interests \cite{whittaker_exclusive_2025}. This threatens public administrations by hindering %
their capacity to deliver citizen-facing public services,  support regulatory agencies and their vetting of ``the validity of industry policy proposals'',  adequately, and more broadly plan nation-wide civil service requirements \cite{shapiro_blowout_2012, wellstead_capacity_2026}. 
Compounding {such} economic risks with the impacts of capture on public values such as safety, health and fundamental rights provides a growing case for governments to reconsider their AI strategy and reflect on how their role in promoting corporate capture may be at odds with the public interest.

\subsubsection{Lessons from adjacent movements}
{
Many of the mechanisms of capture {used} by Big AI, that we have discussed in this paper, mirror strategies that have  historically been {applied} by  similar industries such as Big Tobacco, Big Pharma, and Big Oil.
Civil society's efforts to hold big corporations accountable in these sectors is  ongoing and {has} been met with significant challenges and has often {fallen} short of meaningfully countering corporate power~\cite{gayle2025cop30, american2025state, sanders2024big}. 
Yet, there are remain lesson to be learned from these efforts and the braoder scholarship on corporate capture.
For example, the OECD report
 on preventing policy capture in public decision-making~\cite{oecd2017preventing}  recommends to
\begin{enumerate*}[label=(\roman*)]
\item {level} the playing field by engaging diverse stakeholders,
\item {ensure} transparency and access to information,
\item {promote} accountability via external control, effective competition, and regulatory policies, and
\item {define} clear institutional codes of conduct, {promote} cultures of integrity, and {establish} appropriate frameworks for risk-management.
\end{enumerate*}
{Similarly}, in the context of Big Tobacco,~\citet{lee2025engaging} calls for appropriate separation between public and private interests, binding rules for government-industry interactions to manage conflicts-of-interests, enforcement of transparency and accountability practices, and safeguarding academic knowledge production from undue industry influence. 
A 2024 report~\cite{ainow2024lessons}  further calls for %
applying transferable lessons from the US Food and Drug Administration on how to regulate and hold Big AI accountable.

However, implementing such safeguards %
becomes a challenge when significant corporate capture processes are in progress.
Under these circumstances, activism, collective organizing, and public campaigning can serve to build collective power and pressure policymakers and regulatory bodies to foreground public interests~\cite{rogers2025strategies, thomas2025impacts, harvey2021how}.
For example, \citet{harvey2021how} describe the role of public organising in 2021 during the COVID-19 pandemic to successfully challenge  Big Pharma  and pressure the US government to change its stance on the temporary waiver of the {World Trade Organisation} TRIPS rules to increase global production of COVID-related health technologies.
Similarly, \citet{thomas2025impacts} found strong evidence that climate activism {can shift} public opinion and media representation, and moderate evidence that it can influence voting and how politicians communicate.
Affecting change also requires a mix of tactics spanning protests, lawsuits, lobbying, economic pressure, and coalition-building{. These} collective efforts benefit from identifying critical pressure points, making bolder demands, and strategic coordination between actions~\cite{rogers2025strategies}.

\subsubsection{Counter narratives and resistance to Big AI capture}

Our work on conceptualising and understanding capture supports the growing number of calls and actions to counter corporate capture, dominant narratives, and more broadly, policy agendas that prioritise %
corporate agenda over public interests.
The qualitative insights into the  mechanisms that enable %
capture %
(and to a lesser extent the frequency with which these are reported on) illuminate intervention points for  {resistance} and efforts to promote alternative regulatory interests and counter narratives. These include, but are not limited to:
\begin{enumerate*}
\item\textbf{%
work from civil society organisations
} that advise on regulatory implementation, standard setting, or strategic litigation~\cite{european_center_for_not-for-profit_law_ecnl_towards_2024,BEUCPreserveAIActScope,CDTPreserveAIActScope}, %
\item\textbf{efforts {to} promote shared narratives and bottom-up agendas} for AI policy that are grounded in the experience of members of the public~\cite{ai_now_institute_peoples_2025,PeoplesConsultationCanada},
\item \textbf{work that uncovers the influences that facilitate capture}, in particular regarding lobbying and deregulation\cite{LobbyControl,CorporateEuropeObservatoryLobbying}, 
\item\textbf{independent investigative journalism} that continues to  document and expose numerous issues across the AI supply-chain from corporate power to discriminatory AI systems~\cite{Lighthouse_Reports_2026,bellingcat_2024,team,404Media},
\item\textbf{independent audits} that provide rigorous and reliable %
evidence {of} the workings of AI systems \cite{aif_dataset_2024}, with adequate investment on audit target identification and dissemination of results to support advocacy for fundamental rights \cite{birhane_ai_2024,ojewale_towards_2025},
\item\textbf{efforts that mobilize labour movements to address power asymmetries} ~\cite{merchant_hundreds_2025},  
\item efforts to invigorate existing \textbf{commitments to climate and environmental protection}~\cite{green_screen_coalition_within_2025}, and
\item {work that highlights the} \textbf{necessity to respect rights}  within and outside the boundaries of regulatory concerns~\cite{EDRiRights,ICCLAlgorithm}. Supporting, uplifting, and strengthening {such efforts} is key to {understanding}, challenging, and dismantling corporate capture {while} advancing regulation in the public interest.  
\end{enumerate*}
}

{Finally}, the findings in this paper have implications for scholarship in FAccT and other communities.
The taxonomy provides a scaffolding for other studies and investigations into corporate capture, which can be further refined and enriched. 
It motivates the need to further reflect on the impact of capture on our own scholarship,  what that means for {the community's contributions} to AI regulation, and how we ensure that scientific integrity {and} plurality of perspectives {are} protected in our scholarship and across broader scientific communities and, indirectly, in the associated societies whose institutions lean on our work~\cite{whittaker2021steep}.

\section{Conclusion}
In this study, we have examined the strategies and tactics employed by the Big AI industry to capture regulatory processes. 
Our study does reflects an urgent academic endeavour and a pressing real-world issue with direct implications for people across the globe. 
While democratic regulators ought to attend the concerns of industrial sectors, regulation
should always prioritise protecting and promoting the core public values for which governments bear responsibility. 
The AI industry's power, wealth and influence has far-reaching implications in terms of its impact on the rule of law, the labour market, the environment, knowledge production, and, ultimately, on democratic processes and institutions is so corrosive that policymakers ought to treat it as an emergency. 
Although our findings show the breadth and depth of regulatory capture with no clear immediate path to meaningful accountability and systems of {truly} independent oversight, we hope this work can help lay a foundation for developing and institutionalizing corrective measures. 
This might, for example, take the form of enhancing existing civil society efforts to document and expose the breadth and extent of mechanisms being used, pressuring regulators for increased transparency and accountability in the rule-making processes, and implementing declarations of conflict of interest in regulatory processes in addition to major conference and scientific publication venues in relation to work on societal impacts of AI. 
Government complicity is detrimental  to ensuring the rule of law {and} to restoring trust in public interest technologies. 
Therefore, meaningful regulations and enforcement, developed and enforced in line with the interests of the general public and vulnerable groups, %
is in the interest of  government, regulatory institutions, {and} the AI industry itself.

\section{Limitations}
We have taken all possible measure within our means and control to curate our datasets. Nonetheless, our findings may be influenced by %
sampling bias as well as to changes in reporting style and customs within Reuters. 
Future larger-scale sampling and annotation - by including more events crucial to consensus-building in preparation of regulation, and extending time windows and keywords - may provide an improved quantification of the issues at hand. 
Notwithstanding the large consensus on the high quality of the sources we selected \cite{lin_high_2023}, all publications exist within a larger social and media ecosystem subject to the effects of media capture mentioned in Section \ref{sec:related_work}.
Comparative analysis across geographically and culturally diverse news agencies and publication venues may thus provide improved insight into reporting biases, and afford more refined estimators for the occurrence of both mechanisms of capture and narratives. 
Additionally, such a pluralistic effort may inform further validation and development of our taxonomy of mechanisms of capture. %

\section{Generative AI Usage Statement}

The authors  declare that no form of generative AI was used in the writing of this paper or at any stage of the research process, other than as a support to a senior software engineer in the development of charts. All generated code was thoroughly reviewed line-by-line, edited and tested against its intended specification. 

\section{Acknowledgments}
We would like to thank Matt Davies, %
Petter Ericson, and two experts, a legal scholar from a civil society organisation and an employee at a big corporation, respectively, who prefer to remain anonymous, for their invaluable feedback on this work. We are also grateful for the thorough feedback from the FAccT anonymous reviewers.  
The AI Accountability Lab is supported by grants from the John D. and Catherine T. MacArthur Foundation, the AI Collaborative of the Omidyar Group, Luminate Foundation, European AI \& Society Fund, and Bestseller Foundation. Zeerak Talat is supported by the Arts and Humanities Research Council (grant AH/X007146/1).

\bibliographystyle{ACM-Reference-Format}

\bibliography{CorpCap}

@misc{dehghan_dealing_2025,
	title = {Dealing with {Annotator} {Disagreement} in {Hate} {Speech} {Classification}},
	url = {http://arxiv.org/abs/2502.08266},
	doi = {10.48550/arXiv.2502.08266},
	urldate = {2026-04-25},
	publisher = {arXiv},
	author = {Dehghan, Somaiyeh and Sen, Mehmet Umut and Yanikoglu, Berrin},
	month = aug,
	year = {2025},
	note = {arXiv:2502.08266 [cs]},
}

@online{LetterAICertainty,
    title = {AN OPEN LETTER: Europe needs regulatory certainty on AI},
    author = {Open Letter},
    url = {https://euneedsai.com//},
    year = {2024}
}

@misc{BEUCPreserveAIActScope,
    title = {Open Joint Letter on the Digital Omnibus on AI Preserving the Scope and Integrity of the AI Act},
    date = {2026-04-08},
    author = {European Consumer Organisation (BEUC)},
    howpublished = {\url{https://www.beuc.eu/letters/open-joint-letter-digital-omnibus-ai-preserving-scope-and-integrity-ai-act}}
}

@misc{CDTPreserveAIActScope,
    title = {Joint Open Letter: Preserving the Scope and Integrity of the AI Act},
    date = {2026-04-08},
    author = {Center for Democracy \& Technology},
    howpublished = {https://cdt.org/insights/joint-open-letter-preserving-the-scope-and-integrity-of-the-ai-act/}
}

@misc{EDRiRights,
    title = {The EU must uphold hard-won protections for digital human rights},
    date = {2025-11-13},
    author = {European Digital Rights (EDRi)},
    howpublished = {\url{https://edri.org/wp-content/uploads/2025/11/The-EU-must-uphold-hard-won-protections-for-digital-human-rights.pdf}}
}

@misc{PeoplesConsultationCanada,
    title = {People's Consultation on AI in Canada},
    year = {2026},
    author = {People's Consultation on AI},
    howpublished = {\url{https://www.peoplesaiconsultation.ca/}}
}

@misc{CorporateEuropeObservatoryLobbying,
    title = {This is what corporate capture looks like! Report: How corporations run the EU deregulation agenda },
    date = {2026-04-01},
    author = {Center for Democracy \& Technology},
    howpublished = {\url{https://corporateeurope.org/en/2026/04/what-corporate-capture-looks}}
}

@misc{LobbyControl,
    title = {LobbyFacts - exposing lobbying in the European institutions},
    author = {LobbyFacts},
    year = {2026},
    howpublished = {\url{https://www.lobbyfacts.eu/}}
}

@misc{ICCLAlgorithm,
    date={2023-12-14},
    title = {The European Commission must follow Ireland’s lead, and switch off Big Tech’s toxic algorithms},
    author = {Irish Council for Civil Liberties},
    howpublished={\url{https://www.iccl.ie/2023/the-european-commission-must-follow-irelands-lead-and-switch-off-big-techs-toxic-algorithms/}}
}

@article{afp_concentration_2025,
	title = {Concentration of corporate power a 'huge' concern: {UN} rights chief},
	shorttitle = {Concentration of corporate power a 'huge' concern},
	url = {https://us.afpnews.com/article/?concentration-of-corporate-power-a-huge-concern-un-rights-chief,83MG3CQ},
	language = {en-US},
	urldate = {2026-01-13},
	publisher = {AFP},
	author = {AFP},
	month = nov,
	year = {2025},
}

@article{merchant_hundreds_2025,
	address = {Los Angeles, CA, USA},
	title = {Hundreds of workers mobilize to '{Stop} {Gen} {AI}' and help each other survive {AI} automation},
	url = {https://www.bloodinthemachine.com/p/hundreds-of-workers-mobilize-to-stop},
	language = {en},
	urldate = {2026-01-13},
	journal = {Blood in the Machine},
	author = {Merchant, Brian},
	month = jun,
	year = {2025},
}

@misc{green_screen_coalition_within_2025,
	title = {Within {Bounds}: {Limiting} {AI}'s environmental impact},
	shorttitle = {Within {Bounds}},
	url = {https://greenscreen.network/en/blog/within-bounds-limiting-ai-environmental-impact/},
	language = {en},
	urldate = {2025-12-12},
	journal = {Green Screen Coalition},
	author = {{Green Screen Coalition}},
	month = feb,
	year = {2025},
	note = {Section: blog},
}

@article{the_economist_investors_2025,
	title = {Investors expect {AI} use to soar. {That}’s not happening},
	issn = {0013-0613},
	url = {https://www.economist.com/finance-and-economics/2025/11/26/investors-expect-ai-use-to-soar-thats-not-happening},
	day = 26,
    month = nov,
	year = {2025},
    urldate = {2025-12-02},
	journal = {The Economist},
	author = {The Economist},
}

@misc{arun_bubble_2025,
	title = {Bubble or {Nothing}},
	url = {https://publicenterprise.org/report/bubble-or-nothing/},
	language = {en-US},
	urldate = {2025-12-02},
	journal = {Center for Public Enterprise},
	author = {Arun, Advait},
	month = nov,
	year = {2025},
}

@article{european_commission_simpler_2025,
	address = {Brussels},
	title = {Simpler digital rules to help {EU} businesses grow},
	url = {https://commission.europa.eu/news-and-media/news/simpler-digital-rules-help-eu-businesses-grow-2025-11-19_en},
	language = {en},
	urldate = {2026-01-13},
	author = {{European Commission}},
	month = nov,
	year = {2025},
}

@article{hevner_design_2004,
	title = {Design {Science} in {Information} {Systems} {Research}},
	language = {en},
	journal = {MIS quarterly},
	author = {Hevner, Alan R. and March, Salvatore T. and Park, Jinsoo and Ram, Sudha},
	year = {2004},
	pages = {75--105},
}

@inproceedings{offermann_artifact_2010,
	address = {Berlin, Heidelberg},
	series = {Lecture {Notes} in {Computer} {Science}},
	title = {Artifact {Types} in {Information} {Systems} {Design} {Science} – {A} {Literature} {Review}},
	isbn = {978-3-642-13335-0},
	doi = {10.1007/978-3-642-13335-0_6},
	language = {en},
	booktitle = {Global {Perspectives} on {Design} {Science} {Research}},
	publisher = {Springer},
	author = {Offermann, Philipp and Blom, Sören and Schönherr, Marten and Bub, Udo},
	editor = {Winter, Robert and Zhao, J. Leon and Aier, Stephan},
	year = {2010},
	pages = {77--92},
}

@incollection{vom_brocke_introduction_2020,
	address = {Cham},
	title = {Introduction to {Design} {Science} {Research}},
	isbn = {978-3-030-46781-4},
	url = {https://doi.org/10.1007/978-3-030-46781-4_1},
	language = {en},
	urldate = {2026-01-06},
	booktitle = {Design {Science} {Research}. {Cases}},
	publisher = {Springer International Publishing},
	author = {vom Brocke, Jan and Hevner, Alan and Maedche, Alexander},
	editor = {vom Brocke, Jan and Hevner, Alan and Maedche, Alexander},
	year = {2020},
	doi = {10.1007/978-3-030-46781-4_1},
	pages = {1--13},
}

@misc{corporate_europe_observatory_bias_2025,
	title = {Bias baked in: {How} {Big} {Tech} sets its own {AI} standards},
	url = {https://corporateeurope.org/en/2025/01/bias-baked},
	language = {en},
	urldate = {2026-01-06},
	author = {{Corporate Europe Observatory}},
	month = jan,
	year = {2025},
}

@misc{ai_now_institute_peoples_2025,
	title = {People’s {AI} {Action} {Plan} {Launches} to {Provide} {Counter}-{Weight} to {Trump}’s {Industry}-{Backed} {AI} {Plan} and {EOs}},
	url = {https://ainowinstitute.org/news/announcement/peoples-ai-action-plan-launches-to-provide-counter-weight-to-trumps-industry-backed-ai-plan-and-eos},
	language = {en-US},
	urldate = {2026-01-13},
	journal = {AI Now Institute},
	author = {{AI Now Institute}},
	month = jul,
	year = {2025},
}

@misc{european_center_for_not-for-profit_law_ecnl_towards_2024,
	title = {Towards an {AI} {Act} that serves  people and society: {Strategic} actions for civil society and funders on the enforcement of the {EU} {AI} {Act}},
	url = {https://ecnl.org/sites/default/files/2024-08/241508_AIAct%20implementation_ECNL%20report_final%20design.pdf},
	urldate = {2026-01-13},
	author = {{European Center for Not-for-Profit Law (ECNL)} and {European AI \& Society Fund}},
	month = aug,
	year = {2024},
}

@article{weigand_artifact_2021,
	title = {An artifact ontology for design science research},
	volume = {133},
	journal = {Data \& Knowledge Engineering},
	author = {Weigand, Hans and Johannesson, Paul and Andersson, Birger},
	year = {2021},
	note = {Publisher: Elsevier},
	pages = {101878},
}

@article{hevner_three_2007,
	title = {A three cycle view of design science research},
	volume = {19},
	number = {2},
	journal = {Scandinavian journal of information systems},
	author = {Hevner, Alan R.},
	year = {2007},
	pages = {4},
}

@article{moro2017thalidomide,
  title={The thalidomide tragedy: the struggle for victims' rights and improved pharmaceutical regulation},
  author={Moro, Adriana and Invernizzi, Noela},
  journal={Hist{\'o}ria, Ci{\^e}ncias, Sa{\'u}de-Manguinhos},
  volume={24},
  pages={603--622},
  year={2017},
  publisher={SciELO Brasil}
}

@article{loewenberg2008drug,
  title={Drug company trials come under increasing scrutiny},
  author={Loewenberg, Samuel},
  journal={The Lancet},
  volume={371},
  number={9608},
  pages={191--192},
  year={2008},
  publisher={Elsevier}
}

@article{carr2003pfizer,
  title={Pfizer's epidemic: a need for international regulation of human experimentation in developing countries},
  author={Carr, David M},
  journal={Case W. Res. J. Int'l L.},
  volume={35},
  pages={15},
  year={2003},
  publisher={HeinOnline}
}

@article{dal2006regulatory,
  title={Regulatory capture: A review},
  author={Dal B{\'o}, Ernesto},
  journal={Oxford review of economic policy},
  volume={22},
  number={2},
  pages={203--225},
  year={2006},
  publisher={Oxford University Press}
}

@book{carpenter2013preventing,
  title={Preventing regulatory capture: Special interest influence and how to limit it},
  author={Carpenter, Daniel and Moss, David A},
  year={2013},
  publisher={Cambridge University Press}
}

@article{ahmed2023growing,
  title={The growing influence of industry in AI research},
  author={Ahmed, Nur and Wahed, Muntasir and Thompson, Neil C},
  journal={Science},
  volume={379},
  number={6635},
  pages={884--886},
  year={2023},
  publisher={American Association for the Advancement of Science}
}

@book{levi2011handbook,
  title={Handbook on the Politics of Regulation},
  author={Levi-Faur, David},
  year={2011},
  publisher={Edward Elgar Publishing}
}

@article{shleifer2005understanding,
  title={Understanding regulation.},
  author={Shleifer, Andrei},
  journal={European Financial Management},
  volume={11},
  number={4},
  year={2005}
}

@misc{AIAct,
author={EU},
    title={EU AI Act first regulation on artificial intelligence},
    howpublished={https://www.europarl.europa.eu/topics/en/article/20230601STO93804/eu-ai-act-first-regulation-on-artificial-intelligence },
    year={2023}
}

@incollection{agrell2012rethinking,
  title={Rethinking regulatory capture},
  author={Agrell, Per J. and Gautier, Axel},
  booktitle={Recent advances in the analysis of competition policy and regulation},
  year={2012},
  publisher={Edward Elgar Publishing}
}

@article{li2023regulatory,
  title={Regulatory capture’s third face of power},
  author={Li, Wendy Y.},
  journal={Socio-Economic Review},
  volume={21},
  number={2},
  pages={1217--1245},
  year={2023},
  publisher={Oxford University Press}
}

@inproceedings{wei2024ai,
  title={How Do AI Companies “Fine-Tune” Policy? Examining Regulatory Capture in AI Governance},
  author={Wei, Kevin and Ezell, Carson and Gabrieli, Nick and Deshpande, Chinmay},
  booktitle={Proceedings of the AAAI/ACM Conference on AI, Ethics, and Society},
  volume={7},
  pages={1539--1555},
  year={2024}
}

@article{whittaker2021steep,
  title={The steep cost of capture},
  author={Whittaker, Meredith},
  journal={Interactions},
  volume={28},
  number={6},
  pages={50--55},
  year={2021},
  publisher={ACM New York, NY, USA}
}

@article{lancieri2024ai,
  title={AI Regulation: Competition, Arbitrage \& Regulatory Capture},
  author={Lancieri, Filippo and Edelson, Laura and Bechtold, Stefan},
  journal={Center for Law \& Economics Working Paper Series},
  volume={11},
  year={2024},
  publisher={Center for Law \& Economics, ETH Zurich}
}

@article{ada25,
  title={The Ada Lovelace Institute and The Alan Turing Institute, How do people feel about AI? Wave two of a nationally representative survey of UK attitudes to AI},
  author={The Ada Lovelace Institute and The Alan Turing Institute},
  journal={The Ada Lovelace Institute},
  year={2025}
}

@article{tyson202360,
  title={60\% of Americans would be uncomfortable with provider relying on AI in their own health care},
  author={Tyson, Alec and Pasquini, Giancarlo and Spencer, Alison and Funk, Cary},
  year={2023},
  publisher={Pew Research Center}
}

@inproceedings{abdalla2021grey,
  title={The grey hoodie project: Big tobacco, big tech, and the threat on academic integrity},
  author={Abdalla, Mohamed and Abdalla, Moustafa},
  booktitle={Proceedings of the 2021 AAAI/ACM Conference on AI, Ethics, and Society},
  pages={287--297},
  year={2021}
}

@article{vertinsky2021pharmaceutical,
  title={Pharmaceutical (Re) Capture},
  author={Vertinsky, Liza},
  journal={Yale J. Health Pol'y L. \& Ethics},
  volume={20},
  pages={146},
  year={2021},
  publisher={HeinOnline}
}

@article{savell2014does,
  title={How does the tobacco industry attempt to influence marketing regulations? A systematic review},
  author={Savell, Emily and Gilmore, Anna B and Fooks, Gary},
  journal={PloS one},
  volume={9},
  number={2},
  pages={e87389},
  year={2014},
  publisher={Public Library of Science San Francisco, USA}
}

@article{hiltner2024fossil,
  title={Fossil fuel industry influence in higher education: a review and a research agenda},
  author={Hiltner, Sofia and Eaton, Emily and Healy, Noel and Scerri, Andrew and Stephens, Jennie C and Supran, Geoffrey},
  journal={Wiley Interdisciplinary Reviews: Climate Change},
  volume={15},
  number={6},
  pages={e904},
  year={2024},
  publisher={Wiley Online Library}
}

@book{muttitt2004degrees,
  title={Degrees of capture: Universities, the oil industry and climate change},
  author={Muttitt, Greg},
  year={2004},
  publisher={New Economics Foundation}
}

@misc{noor2024elite,
  title={Elite US universities rake in millions from big oil donations, research finds},
  author={Noor, Dharna },
  year={2024},
  publisher={The Guardian}
}

@misc{taft2024oil,
  title={How Oil Companies Manipulate Journalists},
  author={Taft, Molly},
  year={2024},
  publisher={Drilled}
}

@article{estache2011anti,
  title={Anti-corruption policy in theories of sector regulation},
  author={Estache, Antonio and Wren-Lewis, Liam and Rose-Ackerman, S and Soreide, T},
  journal={Chapter},
  volume={9},
  pages={269--299},
  year={2011}
}

@incollection{krimsky1985corporate,
  title={The corporate capture of academic science and its social costs},
  author={Krimsky, Sheldon},
  booktitle={Genetics and the Law III},
  pages={45--55},
  year={1985},
  publisher={Springer}
}

@article{baba2005legislating,
  title={Legislating “sound science”: the role of the tobacco industry},
  author={Baba, Annamaria and Cook, Daniel M and McGarity, Thomas O and Bero, Lisa A},
  journal={American Journal of Public Health},
  volume={95},
  number={S1},
  pages={S20--S27},
  year={2005},
  publisher={American Public Health Association}
}

@misc{cranor2008tobacco,
  title={The tobacco strategy entrenched},
  author={Cranor, Carl F},
  year={2008},
  publisher={American Association for the Advancement of Science}
}

@article{orr2010merchants,
  title={Merchants of Doubt: How a Handful of Scientists Obscured the Truth on Issues from Tobacco Smoke to Global Warming.},
  author={Orr, David},
  journal={Nature},
  volume={466},
  number={7306},
  pages={565--566},
  year={2010},
  publisher={Nature Publishing Group}
}

@book{krimsky2004science,
  title={Science in the private interest: Has the lure of profits corrupted biomedical research?},
  author={Krimsky, Sheldon},
  year={2004},
  publisher={Bloomsbury Publishing PLC}
}

@article{sass2005vinyl,
  title={Vinyl chloride: a case study of data suppression and misrepresentation},
  author={Sass, Jennifer Beth and Castleman, Barry and Wallinga, David},
  journal={Environmental Health Perspectives},
  volume={113},
  number={7},
  pages={809--812},
  year={2005},
  publisher={National Institue of Environmental Health Sciences}
}

@book{michaels2008doubt,
  title={Doubt is their product: how industry's assault on science threatens your health},
  author={Michaels, David},
  year={2008},
  publisher={Oxford University Press}
}

@article{lachapelle2024academic,
  title={Academic capture in the Anthropocene: a framework to assess climate action in higher education},
  author={Lachapelle, Paul and Belmont, Patrick and Grasso, Marco and McCann, Roslynn and Gouge, Dawn H and Husch, Jerri and de Boer, Cheryl and Molzbichler, Daniela and Klain, Sarah},
  journal={Climatic Change},
  volume={177},
  number={3},
  pages={40},
  year={2024},
  publisher={Springer}
}

@article{hurt2004turning,
  title={Turning free speech into commercial speech: Philip Morris' use of journalists to discredit the EPA report on secondhand smoke},
  author={Hurt, R.D. and Muggli, M.E. and Becker, L.B.},
  journal={Journal of Clinical Oncology},
  volume={22},
  number={14\_suppl},
  pages={6151--6151},
  year={2004},
  publisher={American Society of Clinical Oncology}
}

@misc{schiffrin2018introduction,
  title={Introduction to special issue on media capture},
  author={Schiffrin, Anya},
  journal={Journalism},
  volume={19},
  number={8},
  pages={1033--1042},
  year={2018},
  publisher={SAGE Publications Sage UK: London, England}
}

@incollection{enikolopov2015media,
  title={Media capture: Empirical evidence},
  author={Enikolopov, Ruben and Petrova, Maria},
  booktitle={Handbook of media economics},
  volume={1},
  pages={687--700},
  year={2015},
  publisher={Elsevier}
}

@article{stiglitz2017toward,
  title={Toward a taxonomy of media capture},
  author={Stiglitz, Joseph E},
  year={2017}
}

@article{morgan2019cost,
  title={The cost of capture: how the pharmaceutical industry has corrupted policymakers and harmed patients},
  author={Morgan, Julie Margetta and Duffy, Devin},
  journal={The Roosevelt Institute},
  year={2019}
}

@article{schyns2023lobbying,
  title={The lobbying ghost in the machine},
  author={Schyns, Camille},
  year={2023},
  publisher={Corporate Europe Observatory}
}

@article{gorwa2024platform,
  title={Platform lobbying: Policy influence strategies and the EU's Digital Services Act},
  author={Gorwa, Robert and Lechowski, Grzegorz and Schnei{\ss}, Daniel},
  journal={Internet Policy Review},
  volume={13},
  number={2},
  pages={1--26},
  year={2024},
  publisher={Berlin: Alexander von Humboldt Institute for Internet and Society}
}

@article{lin_high_2023,
	title = {High level of correspondence across different news domain quality rating sets},
	volume = {2},
	issn = {2752-6542},
	url = {https://doi.org/10.1093/pnasnexus/pgad286},
	doi = {10.1093/pnasnexus/pgad286},
	number = {9},
	urldate = {2026-01-13},
	journal = {PNAS Nexus},
	author = {Lin, Hause and Lasser, Jana and Lewandowsky, Stephan and Cole, Rocky and Gully, Andrew and Rand, David G and Pennycook, Gordon},
	month = sep,
	year = {2023},
	pages = {pgad286},
}

@misc{cerulus2025ranked,
  title={Ranked: The 10 most intensely lobbied EU laws},
  author={Cerulus, Laurens and Cokelaere, Hanne and Gros, Marianne and Brzeziński, Bartosz},
  year={2025},
  publisher={Politico}
}

@article{hall2025investing,
  title={Investing in Political Expertise: The Remarkable Scale of Corporate Policy Teams},
  author={Hall, Andrew B and Sun, Anna and Stanford, GSB},
  year={2025}
}

@article{murgia2019ai,
  title={AI academics under pressure to do commercial research},
  author={Murgia, Madhumita},
  journal={Financial Times},
  volume={13},
  year={2019}
}

@inproceedings{young2022confronting,
  title={Confronting power and corporate capture at the FAccT Conference},
  author={Young, Meg and Katell, Michael and Krafft, PM},
  booktitle={Proceedings of the 2022 ACM Conference on Fairness, Accountability, and Transparency},
  pages={1375--1386},
  year={2022}
}

@article{barakat2024selective,
  title={Selective perspectives: A content analysis of The New York Times’ reporting on artificial intelligence},
  author={Barakat, H},
  journal={Computer Says Maybe},
  year={2024}
}

@misc{tanner2023reframing,
  title={Reframing AI in Civil Society: Beyond Risk \& Regulation},
  author={Tanner, Jonathan and Bryden, John},
  year={2023},
  howpublished={\url{https://rootcause.global/framing-ai/}}
}

@misc{robins2025how,
  title={How big tech is creating its own friendly media bubble to ‘win the narrative battle online’},
  author={Robins-Early, Nick},
  year={2025}
}

@misc{de2024inescapable,
  title={Inescapable AI: The Ways AI Decides How Low-Income People Work, Live, Learn, and Survive},
  author={De Liban, Kevin},
  year={2024},
  publisher={Techtonic Justice. https://www.techtonicjustice.org/reports/inescapable-ai}
}

@misc{ainowinstituteArtificialPower,
	author = {AI Now Institute},
	title = {{A}rtificial {P}ower: 2025 {L}andscape {R}eport - {A}{I} {N}ow {I}nstitute --- ainowinstitute.org},
	howpublished = {\url{https://ainowinstitute.org/publications/research/ai-now-2025-landscape-report}},
	year = {2025},
	note = {[Accessed 23-12-2025]},
}

@article{pasquale2024consent,
  title={Consent and compensation: Resolving generative ai's copyright crisis},
  author={Pasquale, Frank and Sun, Haochen},
  journal={Va. L. Rev. Online},
  volume={110},
  pages={207},
  year={2024},
  publisher={HeinOnline}
}

@article{quintais2025generative,
  title={Generative AI, copyright and the AI Act},
  author={Quintais, Jo{\~a}o Pedro},
  journal={Computer Law \& Security Review},
  volume={56},
  pages={106107},
  year={2025},
  publisher={Elsevier}
}

@article{olanipekun2025computational,
  title={Computational propaganda and misinformation: AI technologies as tools of media manipulation},
  author={Olanipekun, Samson Olufemi},
  journal={World Journal of Advanced Research and Reviews},
  volume={25},
  number={1},
  pages={911--923},
  year={2025}
}

@inproceedings{nie2024artificial,
  title={Artificial Intelligence: The Biggest Threat to Democracy Today?},
  author={Nie, Michelle},
  booktitle={Proceedings of the aaai symposium series},
  volume={3},
  number={1},
  pages={376--379},
  year={2024}
}

@misc{europaAntitrustCommission,
	author = {EU},
	title = {{A}ntitrust: {C}ommission fines {G}oogle €4.34 billion for illegal practices regarding {A}ndroid mobile devices to strengthen dominance of {G}oogle\'s search engine --- ec.europa.eu},
	howpublished = {\url{https://ec.europa.eu/commission/presscorner/detail/en/ip_18_4581}},
	year = {2018},
	note = {[Accessed 23-12-2025]},
}

@misc{cnbcAmazonWith,
	author = {Sam Shead},
	title = {{A}mazon hit with \$887 million fine by {E}uropean privacy watchdog  --- cnbc.com},
	howpublished = {\url{https://www.cnbc.com/2021/07/30/amazon-hit-with-fine-by-eu-privacy-watchdog-.html}},
	year = {2021},
	note = {[Accessed 23-12-2025]},
}

@article{solaiman2023evaluating,
  title={Evaluating the social impact of generative ai systems in systems and society},
  author={Solaiman, Irene and Talat, Zeerak and Agnew, William and Ahmad, Lama and Baker, Dylan and Blodgett, Su Lin and Chen, Canyu and Daum{\'e} III, Hal and Dodge, Jesse and Duan, Isabella and others},
  journal={arXiv preprint arXiv:2306.05949},
  year={2023}
}

@inproceedings{dobbe2022system,
  title={System safety and artificial intelligence},
  author={Dobbe, Roel},
  booktitle={Proceedings of the 2022 ACM Conference on Fairness, Accountability, and Transparency},
  pages={1584--1584},
  year={2022}
}

@article{kalluri2025computer,
  title={Computer-vision research powers surveillance technology},
  author={Kalluri, Pratyusha Ria and Agnew, William and Cheng, Myra and Owens, Kentrell and Soldaini, Luca and Birhane, Abeba},
  journal={Nature},
  pages={1--7},
  year={2025},
  publisher={Nature Publishing Group UK London}
}

@book{feldstein2019global,
  title={The global expansion of AI surveillance},
  author={Feldstein, Steven},
  volume={17},
  number={9},
  year={2019},
  publisher={Carnegie Endowment for International Peace Washington, DC}
}

@inproceedings{singh2025epistemic,
  title={Epistemic Destabilization: AI-Driven Knowledge Generation and the Collapse of Validation Systems},
  author={Singh, Bhavneet},
  booktitle={Proceedings of the AAAI/ACM Conference on AI, Ethics, and Society},
  volume={8},
  number={3},
  pages={2387--2398},
  year={2025}
}

@article{hilal2024misinformation,
  title={Misinformation and the demonization of human Rights: the Jordanian Child Rights Law},
  author={Hilal, Ghofran and Hilal, Thawab and Al-Fawareh, Mohammad},
  journal={Cogent Education},
  volume={11},
  number={1},
  pages={2329417},
  year={2024},
  publisher={Taylor \& Francis}
}

@article{robles2025artificial,
  title={Artificial intelligence technology, public trust, and effective governance},
  author={Robles, Pedro and Mallinson, Daniel J},
  journal={Review of Policy Research},
  volume={42},
  number={1},
  pages={11--28},
  year={2025},
  publisher={Wiley Online Library}
}

@misc{corporateeuropeChallengeThee,
	author = {Corporate Europe Observatory},
	title = {{I} {C}hallenge {T}hee --- corporateeurope.org},
	howpublished = {\url{https://corporateeurope.org/en/2023/11/byte-byte}},
	year = {2023},
	note = {[Accessed 26-12-2025]},
}

@misc{corporateeuropetrojan,
	author = {Corporate Europe Observatory},
	title = {{I} {C}hallenge {T}hee --- corporateeurope.org},
	howpublished = {\url{https://corporateeurope.org/en/2024/03/trojan-horses-how-european-startups-teamed-big-tech-gut-ai-act}},
	year = {2024},
	note = {[Accessed 26-12-2025]},
}

@misc{substackRevealedShocking,
	author = {Lucas Amin},
	title = {{R}evealed: “{S}hocking” scale of {B}ig {T}ech’s influence over {L}abour --- democracyforsale.substack.com},
	howpublished = {\url{https://democracyforsale.substack.com/p/revealed-shocking-scale-of-big-tech-influence-labour-peter-kyle-amazon-google-meta}},
	year = {2025},
	note = {[Accessed 26-12-2025]},
}

@misc{metachildsafety,
	author = {Washington Post},
	title = {},
	howpublished = {\url{https://www.washingtonpost.com/investigations/2025/09/08/meta-research-child-safety-virtual-reality/}},
	year = {2025},
	note = {[Accessed 26-12-2025]},
}

@misc{h1bvisa,
	author = {Reuters},
	title = {},
	howpublished = {\url{https://www.reuters.com/world/us/trump-administration-orders-enhanced-vetting-applicants-h-1b-visa-2025-12-04/}},
	year = {2015},
	note = {[Accessed 26-12-2025]},
}

@misc{bbcSocialMedia,
	author = {BBC},
	title = {{U}{K} social media campaigners among five denied {U}{S} visas --- bbc.com},
	howpublished = {\url{https://www.bbc.com/news/articles/cp39kngz008o}},
	year = {2025},
	note = {[Accessed 26-12-2025]},
}

@misc{issueoneTechCozies,
	author = {Issue One},
	title = {{B}ig {T}ech {C}ozies {U}p to {N}ew {A}dministration {A}fter {S}pending {R}ecord {S}ums on {L}obbying {L}ast {Y}ear - {I}ssue {O}ne --- issueone.org},
	howpublished = {\url{https://issueone.org/articles/big-tech-spent-record-sums-on-lobbying-last-year/}},
	year = {2025},
	note = {[Accessed 26-12-2025]},
}

@misc{europaSpeechPresident,
	author = {European Commission},
	title = {{S}peech by {P}resident von der {L}eyen at the {C}openhagen {C}ompetitiveness {S}ummit --- luxembourg.representation.ec.europa.eu},
	howpublished = {\url{https://luxembourg.representation.ec.europa.eu/actualites-et-evenements/actualites/speech-president-von-der-leyen-copenhagen-competitiveness-summit-2025-10-01_en}},
	year = {2025},
	note = {[Accessed 27-12-2025]},
}

@misc{fragdenstaatKoalitionsverhandlungenCDUCSUSPD,
	author = {FragDenStaat},
	title = {{K}oalitionsverhandlungen {C}{D}{U}/{C}{S}{U}/{S}{P}{D} {A}{G} 3 - {D}igitales --- fragdenstaat.de},
	howpublished = {\url{https://fragdenstaat.de/dokumente/258016-koalitionsverhandlungen-cdu-csu-spd-ag-3-digitales/}},
	year = {2025},
	note = {[Accessed 27-12-2025]},
}

@misc{citizenTrumpHalting,
	author = {Public Citizen},
	title = {{H}ow {T}rump {I}s {H}alting {E}nforcement {A}gainst {C}orporate {L}awbreakers --- citizen.org},
	howpublished = {\url{https://www.citizen.org/article/corporate-clemency-trump-enforcement-report/}},
	year = {2025},
	note = {[Accessed 27-12-2025]},
}

@misc{citizenDeletingTech,
	author = {Public Citizen},
	title = {{D}eleting {T}ech {E}nforcement - {P}ublic {C}itizen --- citizen.org},
	howpublished = {\url{https://www.citizen.org/article/deleting-enforcement-trump-big-tech-billion-report/}},
	year = {2025},
	note = {[Accessed 27-12-2025]},
}

@misc{nytimesSiliconValley,
	author = {Eli Tan},
	title = {{S}ilicon {V}alley {P}ledges \$200 {M}illion to {N}ew {P}ro-{A}.{I}. {S}uper {P}{A}{C}s --- nytimes.com},
	howpublished = {\url{https://www.nytimes.com/2025/08/26/technology/silicon-valley-ai-super-pacs.html}},
	year = {2025},
	note = {[Accessed 27-12-2025]},
}

@misc{datacentre,
	author = {Miller, Gabby},
	title = {Data centers have a political problem — and Big Tech wants to fix it},
	howpublished = {\url{https://www.politico.com/news/2025/12/17/data-centers-have-a-political-problem-and-big-tech-wants-to-fix-it-00693695}},
	year = {2025},
	note = {[Accessed 27-12-2025]},
}

@misc{techcrunchSiliconValley,
	author = {Rebecca Bellan},
	title = {{S}ilicon {V}alley is pouring millions into pro-{A}{I} {P}{A}{C}s to sway midterms | {T}ech{C}runch --- techcrunch.com},
	howpublished = {\url{https://techcrunch.com/2025/08/25/silicon-valley-is-pouring-millions-into-pro-ai-pacs-to-sway-midterms/}},
	year = {2025},
	note = {[Accessed 27-12-2025]},
}

@misc{parliamentnewsBritainDelays,
	author = {Parliament Politics Magazine},
	title = {{B}ritain delays {A}{I} regulations to align with {T}rump’s policies --- parliamentnews.co.uk},
	howpublished = {\url{https://parliamentnews.co.uk/britain-delays-ai-regulations-to-align-with-trumps-policies}},
	year = {2025},
	note = {[Accessed 29-12-2025]},
}

@misc{cabrera2025human,
  title={Human rights are universal, not optional: don’t undermine the EU AI act with a faulty code of practice},
  author={Cabrera, L. and Caroli, L. and Harris, D.E.},
  year={2025},
  publisher={Tech Policy Press}
}

@misc{robins2025trump,
  title={Trump signs executive order aimed at preventing states from regulating AI},
  author={Robins-Early, Nick and Kerr, Dara},
  year={2025},
  publisher={The Guardian}
}

@article{baykurt2025gov,
  title={Gov-tech as capture: public infrastructures under data capitalism},
  author={Baykurt, Burcu},
  journal={Information, Communication \& Society},
  pages={1--16},
  year={2025},
  publisher={Taylor \& Francis}
}

@article{bak2025risks,
  title={The Risks of Industry Influence in Tech Research},
  author={Bak-Coleman, Joseph and O'Connor, Cailin and Bergstrom, Carl and West, Jevin},
  journal={arXiv preprint arXiv:2510.19894},
  year={2025}
}

@misc{bettya_building_2024,
	title = {Building {AI} {Technology} for {Europeans} in a {Transparent} and {Responsible} {Way}},
	url = {https://about.fb.com/news/2024/06/building-ai-technology-for-europeans-in-a-transparent-and-responsible-way/},
	language = {en-US},
	urldate = {2026-01-13},
	journal = {Meta Newsroom},
	author = {Meta Newsroom},
	month = jun,
	year = {2024},

    note = {[Accessed 29-12-2025]},
}

@article{pandit2026terms,
  title={Terms of (Ab) Use: An Analysis of GenAI Services},
  author={Pandit, Harshvardhan J. and Blankvoort, Dick A.H. and Luccioni, Sasha and Birhane, Abeba},
  journal={arXiv preprint arXiv:2603.18964},
  year={2026}
}

@article{van2024big,
  title={Big AI: Cloud infrastructure dependence and the industrialisation of artificial intelligence},
  author={Van Der Vlist, Fernando and Helmond, Anne and Ferrari, Fabian},
  journal={Big Data \& Society},
  volume={11},
  number={1},
  pages={20539517241232630},
  year={2024},
  publisher={SAGE Publications Sage UK: London, England}
}

@misc{gayle2025cop30,
  title={Cop30 was meant to be a turning point, so why do some say the climate summit is broken?},
  author={Gayle, Damien},
  year={2025},
  publisher={The Guardian}
}

@misc{american2025state,
  title={``State of Tobacco Control'' 2025: Tobacco Industry’s Aggressive Actions to Protect its Profits Slows Proven Policies to Prevent and Reduce Tobacco Use},
  author={{The American Lung Association}},
  year={2025},
  url={https://www.lung.org/content/sotc/2025/ala-state-of-tobacco-control-2025.pdf}
}

@misc{sanders2024big,
  title={Big Pharma’s Business Model: Corporate Greed},
  author={{Health, Education, Labor, and Pensions Committee (Chair: Bernard Sanders)}},
  year={2024},
  url={https://www.help.senate.gov/imo/media/doc/big_pharmas_business_model_report.pdf}
}

@misc{rogers2025strategies,
  title={Strategies and Tactics to Curb the Fossil Fuel Industry},
  author={Rogers, Cathy and Ostarek, Markus and Nadel, Sam},
  year={2025},
  url={https://www.socialchangelab.org/tactics-curb-fossil-fuel-corporations}
}

@article{thomas2025impacts,
  title={The impacts of climate activism},
  author={Thomas-Walters, Laura and Scheuch, Eric G and Ong, Abby and Goldberg, Matthew H},
  journal={Current Opinion in Behavioral Sciences},
  volume={63},
  pages={101498},
  year={2025},
  publisher={Elsevier}
}

@misc{lee2025engaging,
  title={Engaging policymakers on the Commercial Determinants of Health: Lessons from Global Tobacco Control},
  author={Lee, Kelley},
  year={2025},
  url={https://www.publichealthontario.ca/-/media/Event-Presentations/25/07/engaging-policymakers-global-tobacco-control.pdf}
}

@article{harvey2021how,
  title={How to Fight Big Pharma --- and Win },
  author={Harvey, Ryan and  Foley, Melanie},
  year={2021},
  publisher={Truthout},
  url={https://truthout.org/articles/how-to-fight-big-pharma-and-win/}
}

@book{oecd2017preventing,
  title={Preventing policy capture: Integrity in public decision making},
  author={OECD},
  year={2017},
  publisher={OECD Publications Centre}
}

@article{shapiro_blowout_2012,
	title = {The Complexity of Regulatory Capture: Diagnosis, Causality and Remediation},
	volume = {17},
	issn = {1090-3968},
	shorttitle = {Blowout},
	number = {1},
	journal = {Roger Williams University Law Review},
	author = {Shapiro, Sidney},
	date = {2012-01-12},

}

@inproceedings{oortwin2021,
  title = {Interrater {{Disagreement Resolution}}: {{A Systematic Procedure}} to {{Reach Consensus}} in {{Annotation Tasks}}},
  shorttitle = {Interrater {{Disagreement Resolution}}},
  booktitle = {Proceedings of the {{Workshop}} on {{Human Evaluation}} of {{NLP Systems}} ({{HumEval}})},
  author = {Oortwijn, Yvette and Ossenkoppele, Thijs and Betti, Arianna},
  editor = {Belz, Anya and Agarwal, Shubham and Graham, Yvette and Reiter, Ehud and Shimorina, Anastasia},
  year = 2021,
  month = apr,
  pages = {131--141},
  publisher = {Association for Computational Linguistics},
  address = {Online},
  urldate = {2026-04-05},
  }

@online{ExclusiveMistralSeeks,
  title = {Exclusive: {{Mistral}} Seeks Defence Contracts across {{Europe}}},
  howpublished = {\url{https://sifted.eu/articles/mistral-helsing-defence-ai-action-summit-paris/}},
  date = {2026-04-16},
  author = {Sifted}
}

@misc{ainow2024lessons,
	title = {Lessons from the FDA for AI},
	url = {https://ainowinstitute.org/publications/research/lessons-from-the-fda-for-ai},
	author = {{AI Now Institute}},
	year = {2024}
}

@book{mcquillan_resisting_2022,
	location = {Bristol, {UK}},
	title = {Resisting {AI}: an anti-fascist approach to artificial intelligence},
	isbn = {978-1-5292-1349-2 978-1-5292-1350-8},
	shorttitle = {Resisting {AI}},
	pagetotal = {181},
	publisher = {Bristol University Press},
	author = {{McQuillan}, Dan},
	date = {2022},

}

@inproceedings{birhane_ai_2024,
	location = {Toronto, {ON}, Canada},
	title = {{AI} auditing: The Broken Bus on the Road to {AI} Accountability},
	rights = {https://doi.org/10.15223/policy-029},
	isbn = {979-8-3503-4950-4},
	url = {https://ieeexplore.ieee.org/document/10516659/},
	doi = {10.1109/SaTML59370.2024.00037},
	shorttitle = {{AI} auditing},
	eventtitle = {2024 {IEEE} Conference on Secure and Trustworthy Machine Learning ({SaTML})},
	pages = {612--643},
	booktitle = {2024 {IEEE} Conference on Secure and Trustworthy Machine Learning ({SaTML})},
	publisher = {{IEEE}},
	author = {Birhane, Abeba and Steed, Ryan and Ojewale, Victor and Vecchione, Briana and Raji, Inioluwa Deborah},
	urldate = {2025-11-27},
	date = {2024-04-09},
}

@inproceedings{ojewale_towards_2025,
	location = {Yokohama Japan},
	title = {Towards {AI} Accountability Infrastructure: Gaps and Opportunities in {AI} Audit Tooling},
	isbn = {979-8-4007-1394-1},
	url = {https://dl.acm.org/doi/10.1145/3706598.3713301},
	doi = {10.1145/3706598.3713301},
	shorttitle = {Towards {AI} Accountability Infrastructure},
	eventtitle = {{CHI} 2025: {CHI} Conference on Human Factors in Computing Systems},
	pages = {1--29},
	booktitle = {Proceedings of the 2025 {CHI} Conference on Human Factors in Computing Systems},
	publisher = {{ACM}},
	author = {Ojewale, Victor and Steed, Ryan and Vecchione, Briana and Birhane, Abeba and Raji, Inioluwa Deborah},
	urldate = {2025-11-11},
	date = {2025-04-26},
	langid = {english},
}

@article{aif_dataset_2024,
	title = {A Dataset to Assess Microsoft Copilot Answers in the Context of Swiss, Bavarian and Hessian Elections},
	volume = {18},
	rights = {Copyright (c) 2024 Association for the Advancement of Artificial Intelligence},
	issn = {2334-0770},
	url = {https://ojs.aaai.org/index.php/ICWSM/article/view/31446},
	doi = {10.1609/icwsm.v18i1.31446},

	pages = {2040--2050},
	journaltitle = {Proceedings of the International {AAAI} Conference on Web and Social Media},
	author = {Romano, Salvatore and Angius, Riccardo and Kerby, Natalie and Bouchaud, Paul and Amidei, Jacopo and Kaltenbrunner, Andreas},
	urldate = {2026-04-19},
	date = {2024-05-28},
	langid = {english},
}

@online{whittaker_exclusive_2025,
	title = {Exclusive: Elon Musk staffer created a {DOGE} {AI} assistant for making government “less dumb”},
	url = {https://techcrunch.com/2025/02/18/elon-musk-staffer-created-a-doge-ai-assistant-for-making-government-less-dumb/},
	shorttitle = {Exclusive},
	titleaddon = {{TechCrunch}},
	author = {Whittaker, Zack, Charles Rollet},
	urldate = {2026-04-20},
	date = {2025-02-18},
	langid = {american},
}

@article{wellstead_capacity_2026,
	title = {Capacity for what? Elon Musk’s {DOGE}, public value destruction and the darkside of policy capacity},
	issn = {1727-2645},
	url = {https://doi.org/10.1108/PAP-12-2024-0200},
	doi = {10.1108/PAP-12-2024-0200},
	shorttitle = {Capacity for what?},
	pages = {1--14},
	journaltitle = {Public Administration and Policy},
	shortjournal = {Public Administration and Policy},
	author = {Wellstead, Adam and Howlett, Michael},
	urldate = {2026-04-20},
	date = {2026-04-01},
}

@online{Lighthouse_Reports_2026, 
    title = {Lighthouse Reports},
    author={Lighthouse Reports},
    url={https://www.lighthousereports.com/about/}, 
    year={2026}
}

@misc{bellingcat_2024, 
    url={https://www.bellingcat.com/},
    author={Bellingcat}, 
    year={2026}, month={Mar}}

@misc{team, year={2026}, title={The Nerve}, url={https://www.thenerve.news/}, journal={The Nerve}, author={Cadwalladr, Carole}}

@online{404Media,
    author = {404 Media},
    year = {2026},
    url = {https://www.404media.co/},
    title = {404 Media}
}

\appendix
\section{Taxonomy and Annotation Template}
\subsection{Taxonomy of Mechanisms of Capture and corresponding descriptions}
\label{appendix:taxonomy}

\begin{enumerate}
\item \textbf{Direct influence on policy:} Mechanisms whose aim is to influence the position or decisions of public officials and regulations
\begin{enumerate}
\item \textit{Lobbying: }The communication made by a person or organisation regarding themselves or on behalf on another entity to a public official a stakeholder with the goal of influencing decisions or creating a favourable position for themselves
\item \textit{Private meetings with regulators (outside lobbying):} Meetings and communications that occur directly between a person or organisation and public officials outside of channels regulated under lobbying transparency laws
\item \textit{Political contributions:} Contributions by a person or organisation to a political entity or person
\item \textit{Economic coercion of government:} The use of potential economic advantage or disadvantage as a threat to induce changes in decisions or positions of the government
\end{enumerate}
\item \textbf{Conflicting involvement:} Mechanisms that represent inherent conflicts due to the involvement of an entity in a specific position
\begin{enumerate}
\item \textit{Revolving door:} The pattern of governmental or public officials taking up positions or roles in private entities in areas which they regulated or had authority over, as well as the inverse where public officials are appointed from private entities that were the subject of enforcement and regulation
\item \textit{Direct involvement in rule-making:} Involvement of private industry, governmental, or other entities in the process of developing law or policy without a clearly defined legal mandate or authorisation to do so
\item \textit{Ownership/Stake in company:} Public officials taking an ownership or other forms of stocks or stake in an organisation that is being regulated or overseen by the office they are a part of
\end{enumerate}
\item \textbf{Market influence:} Mechanisms that involve participation, co-operation, or coercion of market actors
\begin{enumerate}
\item \textit{Standard setting in consortia:}Utilising favourable consortiums to develop, set, and dictate standards or pactices without the involvement of other relevant stakeholders
\item \textit{Corralling SMEs/orgs to oppose:}  Imploring, organising, or using small and medium-sized enterprises (SMEs) or their representative organisations to oppose a policy or position
\item \textit{Standard setting via monopoly:} Using a position of monopoly or dominance to dictate practices
\item \textit{Economic coercion of competition: }The use of potential economic advantage or disadvantage as a threat to induce changes in decisions or practices of competitors
\end{enumerate}
\item \textbf{Elusion of law}: Mechanisms that are directly or indirectly against the spirit or the letter of the law 
\begin{enumerate}
\item \textit{Disregard existing laws}: Practices that disregard the requirements or process established in laws
\item \textit{Misinterpret laws}:Using an alternative or intentional misrepresentation of the law as a way to deflect or avoid its requirements or implications
\item \textit{Relocation of development and labour}: Moving labour deployments or setups from one location to another
\item \textit{Exploit weak regulations/jurisdictions}:  Moving organisations, procurements, or deployments to jurisdictions that have considerably weaker regulations or enforcement
\item \textit{Retaliation against whistleblowers \& authorities}:  Acting against whistleblowers or authorities directly or indirectly, or taking actions that lead to negative consequences or suppress the voice or actions of whistle blowers
\item \textit{Bribery:} The solicitation, payment, or use of favour in exchange for official actions or decisions
\end{enumerate}
\item \textbf{Epistemic \& discourse influence:} Mechanisms that create or use narratives as a way to promote a specific practice or opinion
\begin{enumerate}
\item \textit{Corporate sponsorship of events:} Utilising sponsorship of events to promote the organisation or imply its support to specific principles or topics
\item \textit{Funding and sponsorship of research \& education:} Direct or indirect funding and sponsorship of research conducted outside the organisation, typically in educational and academic sectors
\item \textit{Public facing campaign:}  Use of media or traditional media or other forms of public facing media to advertise, promote, or create a narrative against a specific person, topic, or action and/or promote a positive and misleading narrative about a company, AI product or practice 
\item \textit{Hyping technologies:  }Promoting or implying capabilities of technologies on the basis of hypothetical speculations and without empirical evidence or theoretical and conceptual rigour
\item \textit{Playing victim:} Industry representatives complaining via public campaign, directly to regulators or via other means that they have been subject to unfair rules 
\item \textit{Undermining risks/harms:} Disregarding, diminishing, hiding, or otherwise reducing harms or the potential for risks or harms associated with AI technologies
\item \textit{Speculative studies:} Use of studies, analysis, or reports that do not follow standard scientific or conceptual rigour %
\item \textit{Ethics washing:} Companies or corporations imposing %
as the guardians of ethics, transparency, accountability, and safety without meaningful actions or implementations to support the claims or while engaging in practices that undermine these principles 
\item \textit{Government adopting industry framing:}  Governments, public bodies, or public officials adopting, repeating, or only considering framings, positions, or approaches favoured by industry while ignoring, refusing, or diminishing other approaches
\item \textit{Conflation of public and private interest:} Irrationally mixing or implying private interests as being public benefits even though they specifically benefit non-public actors without a corresponding benefit to the public
\end{enumerate}
\end{enumerate}

\subsection{Annotation template}
\label{appendix:codebook}

\begin{enumerate}
\item \textit{Article ID:} a unique identifier for the article
\item \textit{Title:} the title as used in the article
\item \textit{Published:} date of publication of article
\item \textit{Annotator:} the person who performed the annotation
\item \textit{Duplicate of:} ID of article for which this is a duplicate
\item \textit{Mechanism Category:} applicable mechanisms from the taxonomy (see Appendix.\ref{appendix:taxonomy}. If none is applicable, the additionally provided option \textit{No capture} is selected
\item \textit{Mechanism:} description of the mechanism as it appears or is evident in the article
\item \textit{Evidence:} excerpts from the article text which exhibit or describe the mechanism or its effects
\item \textit{Narratives used:} identified narratives from the article
\item \textit{Notes:} space for annotators to capture relevant information, perspectives, or questions for discussion
\item \textit{Consensus:} mechanisms selected after discussion between pair of annotators in subsequent stages
\item \textit{Out of scope:} indicator to signal this article is out of scope
\end{enumerate}

\end{document}